\documentclass[10pt,journal,a4paper]{IEEEtran}
\usepackage {textcomp, gensymb}
\usepackage{amssymb}
\usepackage{amsmath}
\usepackage{bm}
\usepackage{comment}
\usepackage{amsthm}
\usepackage{graphicx}
\usepackage{gensymb}
\usepackage{subfigure}
\usepackage{float}
\usepackage{flushend}
\usepackage{cite}
\usepackage{algorithm,algorithmicx}
\usepackage{algpseudocode}
\algtext*{EndIf}
\usepackage{setspace}
\usepackage{enumerate}
\usepackage{enumitem}
\usepackage{multirow}
\usepackage{amsfonts}
\usepackage{url}
\usepackage{diagbox}
\usepackage{color, soul}
\usepackage{amsthm}
\usepackage{type1cm}

\algnewcommand\algorithmicinput{\textbf{INPUT: }}
\algnewcommand\Input{\item[\algorithmicinput]}
\algnewcommand\algorithmicoutput{\textbf{OUTPUT: }}
\algnewcommand\Output{\item[\algorithmicoutput]}

\newtheorem{proposition}{\bf~~Proposition}

\newtheorem{lemma}{\bf~~Lemma}
\newtheorem{observation}{\bf~~Observation}

\begin{document}
\title{Hybrid Near-Far Field Channel Estimation for Holographic MIMO Communications}

\author{
\IEEEauthorblockN{
    {Shaohua Yue},
	{Shuhao Zeng}, \IEEEmembership{Student Member, IEEE},
	{Liang Liu}, \IEEEmembership{Member, IEEE},
    {Yonina C. Eldar}, \IEEEmembership{Fellow, IEEE},
	{and Boya Di}, \IEEEmembership{Member, IEEE}}
	\vspace{-3mm}
    
    \thanks{Shaohua Yue, Shuhao Zeng, and Boya Di are with the State Key Laboratory of Advanced Optical Communication Systems and Networks, School of Electronics, Peking University, Beijing 100871, China. (e-mail: \{yueshaohua; shuhao.zeng; boya.di\}@pku.edu.cn).} 
    \thanks{Liang Liu is with the Department of Electronic and Information Engineering, The Hong Kong Polytechnic University, Hong Kong, SAR, China. (e-mail: liang-eie.liu@polyu.edu.hk).}
    \thanks{Yonina C. Eldar is with the Faculty of Mathematics and Computer Science, Weizmann Institute of Science, Rehovot 7610001, Israel (e-mail: yonina.eldar@weizmann.ac.il).}
    \thanks{Part of this work has been accepted for publication in the IEEE GLOBECOM 2023 conference\cite{conf}.}
}
\maketitle

\begin{abstract}

Holographic MIMO communications, enabled by large-scale antenna arrays with quasi-continuous apertures, is a potential technology for spectrum efficiency improvement. 
However, the increased antenna aperture size extends the range of the Fresnel region, leading to a hybrid near-far field communication mode. 
The users and scatterers randomly lie in near-field and far-field zones, and thus, conventional far-field-only and near-field-only channel estimation methods may not work. 
To tackle this challenge, we demonstrate the existence of the \emph{power diffusion} (PD) effect, which leads to a mismatch between the hybrid-field channel and existing channel estimation methods. 
Specifically, in far-field and near-field transform domains, the power gain of one channel path may diffuse to other positions, thus generating fake paths. This renders the conventional techniques unable to detect those real paths. 
We propose a PD-aware orthogonal matching pursuit algorithm to eliminate the influence of the PD effect by identifying the PD range, within which paths diffuse to other positions. 
PD-OMP fits a general case without prior knowledge of near-field and far-field path numbers and the user’s location. 
The computational complexity of PD-OMP and the Cram\'er-Rao Lower Bound for the sparse-signal-recovery-based channel estimation are also derived. 
Simulation results show that PD-OMP outperforms state-of-the-art hybrid-field channel estimation methods.

\end{abstract}

 \section{Introduction}\label{sec:intro}
To fulfill the high spectrum efficiency requirement of the future sixth-generation (6G) network~\cite{6G}, holographic MIMO communication has been proposed as a promising solution\cite{hcommun,hdma}, where numerous antenna elements are integrated into a compact two-dimensional surface~\cite{holocom}. 
Potential implementation technologies include reconfigurable holographic surface~\cite{RHS, RHS2} and extremely large reconfigurable intelligent surface~\cite{LIS,lis2,dbyhbf}. 
Due to the increased radiation aperture size of the antenna array, the Fresnel region (radiating near-field region of the antenna) is significantly enlarged~\cite{extendnear}. As a result, part of the users and scatterers lie in the near-field region of the holographic antenna array~\cite{nfs}, where the electromagnetic (EM) waves are characterized by spherical waves~\cite{sphericalwave}. The remaining users and scatterers are located in the far-field region and the EM waves can be modeled via uniform plane waves. This gives rise to the so-called \emph{hybrid near-far field} communication~\cite{hfomp,hfmodel}. 


Due to the modeling difference between near-field and far-field EM wave propagation, conventional channel estimation methods cannot be directly applied to the hybrid-field case, necessitating the development of new schemes. Most existing works focus on either near-field channel estimation~\cite{Polar Domain,nfce,ynearce} or far-field channel estimation~\cite{classicomp}. In \cite{classicomp, Polar Domain}, the polar domain and angular domain channel representation are proposed, respectively, to depict the characteristics of the near-field and far-field channel models. 
In \cite{nfce}, the near-field channel estimation problem is investigated, where the near-field region is divided into grids to perform on-grid estimation. A new dictionary is designed for near-field sparse channel representation and estimation in \cite{ynearce} to relieve the high coherence burden of the dictionary because of the two-dimensional near-field channel representation.
Some initial works \cite{hfomp,sdomp} consider the concept of a hybrid-field channel. Channel estimation techniques are designed relying on prior knowledge of the number of near-field and far-field paths such that the near-field and far-field path components are estimated separately \cite{hfomp,sdomp}. 

However, there is a \emph{power diffusion effect} in the sparse-signal-recovery-based hybrid-field channel estimation, which has not been discovered in the existing literature. To be specific, due to the high coherence between certain near-field and far-field steering vectors, when a near-field (or far-field) path component is transformed from the spatial domain into the angular domain (polar domain), the power gain of this path component spreads to multiple steering vectors and generates fake paths. Such an effect leads to an inaccurate path component estimation in the transform domain, which refers to the near-field polar domain, far-field angular domain, and the hybrid-field joint angular-polar domain, i.e., a concatenation of the polar domain and angular domain. This inaccuracy consequently causes estimation errors for the spatial-domain hybrid-field channel. Moreover, for the general case where the number of near-field and far-field paths are unknown, existing hybrid-field channel estimation algorithms ~\cite{hfomp,sdomp} are not applicable.

In this paper, we investigate hybrid near-far field channel estimation without any prior knowledge of the numbers of near-field and far-field paths. Against this background, two new challenges arise. \emph{First}, it is non-trivial to distinguish the far-field and the near-field paths in the hybrid-field case. The boundary of the near-field and far-field regions is hard to specify since it changes with the propagation direction of the EM wave~\cite{channelmodel}. \emph{Second}, due to the power diffusion effect, the power gain of near-field paths and far-field paths is coupled. It leads to inaccurate transform-domain path component estimation, which urges efficient channel estimation schemes. 
 
To cope with the above challenges, we develop a power-diffusion-aware orthogonal matching pursuit algorithm (PD-OMP) for hybrid near-far field multipath channel estimation. 
The key idea of PD-OMP is two-fold. 
\emph{First}, we observe that the angular domain can only provide accurate information about far-field path components while the polar domain only provides accurate information about near-field path components. Thus, we transform the spatial-domain hybrid-field channel to a joint angular-polar domain, where the near-field and far-field path components are successfully separated. 
\emph{Second}, we define the \emph{power diffusion range} to quantify the power gain diffusion from each path component to other positions. We demonstrate that the power diffusion range of each path is positively related to its power gain. Hence, by estimating the power gain, direction, and propagation distance of each path component, the power diffusion range is calculated so that the interference brought by the power diffusion effect can be identified and eliminated. In this way, the information on each path is extracted regardless of the power diffusion effect, and the hybrid-field channel is estimated accurately. 

Our contributions are summarized below.
\begin{enumerate}
    \item We analyze the power diffusion effect in the sparse-signal-recovery-based hybrid-field channel estimation, which leads to inaccurate transform-domain path component estimation. It indicates that when a multipath hybrid-field channel is transformed from the spatial domain into the angular domain or the polar domain, the power gain corresponding to a path spreads to other positions and generates fake paths. This reveals why conventional far-field and near-field sparse-signal-recovery-based channel estimation methods cannot be directly applied to the hybrid-field case. The power diffusion effect is also demonstrated by an illustrative example and quantified by the \emph{power diffusion range.}
    \item We develop a power-diffusion-aware hybrid-field channel estimation method (PD-OMP), which does not require any prior information on the number of far-field and near-field paths to perform channel estimation. The joint angular-polar domain channel transform is utilized in PD-OMP so that different path components of the multipath channel are separated. Moreover, employing an iterative compressed sensing-based method, PD-OMP introduces the power diffusion range to resolve the inaccurate transform-domain path component estimation, which is different from other hybrid-field channel estimation techniques. 
    \item Simulation results show that the proposed PD-OMP achieves a higher estimation accuracy than current hybrid-field channel estimation techniques, which do not consider the power diffusion effect, given different SNRs and pilot lengths. The influence of scatterer distribution and the considered power diffusion range on the algorithm performance are also discussed. 
\end{enumerate}

The rest of this paper is organized as follows. In section~\ref{sec:sys_mod}, the holographic MIMO communication scenario, the hybrid-field channel model, and the signal model are described. In Section~\ref{sec:transform}, we present the hybrid-field channel characteristics in the joint angular-polar domain and analyze the power diffusion effect. A hybrid-field channel estimation method PD-OMP is proposed and the Cram\'er-Rao Lower Bound of the sparse-signal-recovery-based hybrid-field channel estimation is derived in Section~\ref{sec:algorithm}. In Section~\ref{sec:simulation_results}, simulation results are provided and conclusions are drawn in Section~\ref{sec:conclusion}.

Throughout the paper, we use the following notation.
Vectors and matrices are represented by lower-case and upper-case boldface letters, respectively. The writing $\mathbf{X} \in \mathbb{C}^{a \times b}$ means that the size of $\mathbf{X}$ is ${a \times b}$ and each element of $\mathbf{X}$ is a complex number. In addition, $\mathbf{X}(p,:)$ and $\mathbf{X}(:,p)$ denote the $p$-th row and $p$-th column of the matrix $\mathbf{X}$, respectively. We use $(\cdot)^T, (\cdot)^H$ and $(\cdot)^\dagger$
to denote the transpose, conjugate transpose, and pseudo-inverse operation respectively.  $|\cdot|$ is the absolute operator, ${\rm Tr}(\cdot)$ is the trace operator, and $\mathbb{B}(\mathbf{X}_1, ..., \mathbf{X}_N)$ represents a block diagonal matrix generated from matrices $\mathbf{X}_1, ..., \mathbf{X}_N$. ${\rm card}(\mathbb{A})$ denotes the number of elements in set $\mathbb{A}$.
\section{System Model}\label{sec:sys_mod}
In this section, we first describe the holographic MIMO communication scenario. The far-field and near-field path modeling are then given, respectively, based on which the hybrid-field channel model is presented. The signal model for pilot signal transmission is also provided.
\subsection{Scenario Description}
\begin{figure}[ht]
	\centerline{\includegraphics[width=9cm,height=5.4cm]{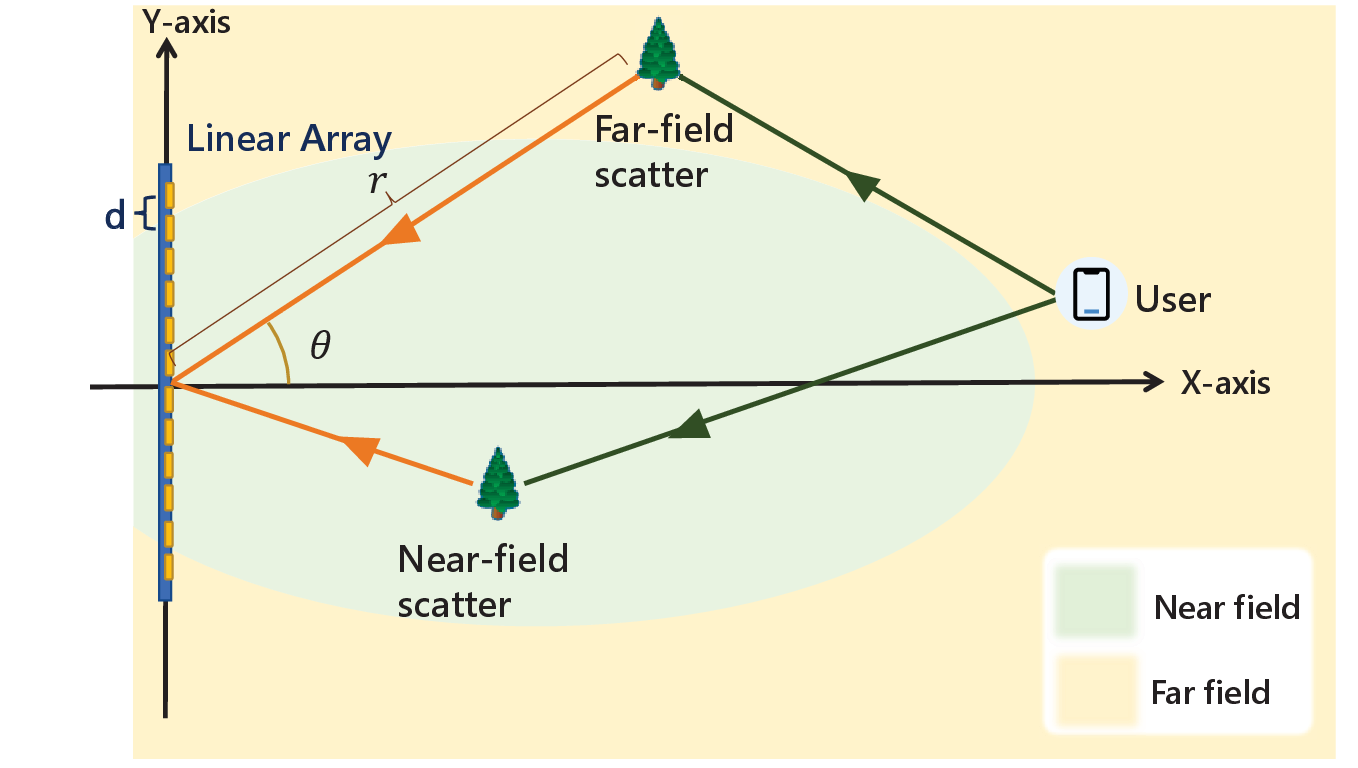}}
	\caption{Hybrid channel model with a linear holographic antenna array.}
	\label{scenario}
\end{figure}

As shown in Fig.~\ref{scenario}, we consider an uplink communication system. The base station (BS) is equipped with an extremely large linear antenna array\footnote{The discussion on channel estimation in this paper is also applicable to the case of a planar array.} to communicate with a single-antenna user, with the number of antenna elements and the element spacing denoted by $N$ and $d$, respectively. We assume that the antenna elements are connected via $N_{RF}<N$ radio frequency (RF) chains such that the analog precoding scheme is employed at the BS.
The EM radiation field of the antenna array can be divided into the near field and far field, as indicated in Fig.~\ref{scenario}. The boundary between these two fields depends on Rayleigh distance, which is positively correlated with the size of the antenna array \cite{channelmodel}. Given the large size of the holographic antenna array, the near-field region extends, leading to \emph{hybrid-field} communications, i.e., the user and scatterers can be located in either the near field or the far field of the antenna array.


\subsection{Hybrid-field Channel Model}
Assume that the hybrid-field multipath channel from the user to the antenna array at the BS consists of a line-of-sight (LoS) path, denoted by path $0$, and $L-1$ non-line-of-sight (NLoS) paths, denoted by paths $1, 2,..., L-1$. 
In the following, we refer to the LoS path as far-field (or near-field) if the user lies in the far-field (or near-field) region of the antenna array and we refer to an NLoS path as far-field (or near-field) if the scatterer corresponding to this path locates in the far-field (or near-field) region of the antenna array.
For the hybrid-field channel, the $L$ paths consist of both far-field and near-field paths. We first present the model for the far-field and near-field paths, respectively, which are then combined to obtain the hybrid-field channel.
For simplicity, we introduce a Cartesian coordinate system, where the $x$-axis is perpendicular to the linear antenna array and the $y$-axis is aligned with the antenna array. The location of the middle point of the antenna array is set to be $(0,0)$, as depicted in Fig.~\ref{scenario}.

\subsubsection{Far-field Path Modeling}
For the user or the scatterer located in the far field of the antenna array, the EM wave of the far-field path received by the antenna array can be approximated by a uniform plane wave. 
In this case, if path $l$ is a far-field path, where $l$ is the index of path based on (\ref{hfchannelmodel}), the model for this path is described as~\cite{nfmodel}
\begin{equation}
	\mathbf{h}_{F,l} = g_l \mathbf{a}(\theta_l),\label{ffchannelmodel}
\end{equation}
where for the LoS path, i.e., $l=0$, $g_l $ represents the channel fading and $\theta_l$ is the angle between the $x$-axis and the direction from the origin to the user. For the NLoS path, i.e., $l \geq 1$, $g_l$ is a random complex factor describing the joint impact of scattering and channel fading, and $\theta_l$ is the angle between the $x$-axis and the direction from the origin to the $l$-th scatterer. $\mathbf{a}(\theta_l)$ represents the far-field steering vector toward $\theta_l$, i.e.,
\begin{equation}
\mathbf{a}(\theta_l)= \frac{1}{\sqrt{N}}[1, e^{j\frac{2\pi d}{\lambda} \sin(\theta_l)}, ..., e^{j\pi \frac{2\pi(N-1)d}{\lambda} \sin(\theta_l)}]^{T}.\label{ffsv}
\end{equation}

\subsubsection{Near-field Path Modeling}
When the user or the scatterer is located in the near field of the antenna array, we use a spherical wave model to describe the wavefront of EM waves, which is more accurate than the plane wave model. To capture this feature, 
if path $l$ is a near-field path, the model for this path is described as~\cite{nfmodel}
\begin{equation}
	\mathbf{h}_{N,l} = g_l \mathbf{b}(\theta_l,r_l),
\end{equation}
where for the LoS path, i.e., $l = 0$, $r_l$ is the distance between the origin and the user. For the NLoS path, i.e.,  $l \geq 1$, $r_l$ is the distance between the origin and the $l$-th scatterer. The term $\mathbf{b}(\theta_l,r_l)$ is the near-field steering vector, expressed as
\begin{equation}
	\mathbf{b}(\theta_l,r_l) = \frac{1}{\sqrt{N}}{[e^{-j\frac{2\pi}{\lambda}(r_{1,l}-r_l)},...,e^{-j\frac{2\pi}{\lambda}(r_{N,l}-r_l)}]}^T,\label{nfsv}
\end{equation}
where $r_{n,l}$ is the distance between the $n$-th antenna element of the antenna array and the user or the $l$-th scatterer corresponding to this path. The term $r_{n,l}$ can be written as
\begin{equation}
	r_{n,l}=\sqrt{(r_l \cos\theta_l)^2+(t_n d-r_l\sin\theta_l)^2},\label{distance}
\end{equation}
where $t_n = \frac{2n-N+1}{2}$ and $(0, t_nd)$ is the coordinate of the $n$-th antenna element.

\subsubsection{Overall Hybrid-field Channel Modeling}
Among the $L$ paths, the set of the far-field and near-field paths from the user to the antenna array is denoted by $\mathbb{L}_F$ and $\mathbb{L}_N$, respectively, i.e., $L = {\rm card}(\mathbb{L}_F) + {\rm card}(\mathbb{L}_N).$ By combining  near-field path components and far-field components, the hybrid-field multipath channel  from the user to the BS is modeled as
\begin{equation}
	\mathbf{h}_{H} = \sum_{l \in \mathbb{L}_{F}} \mathbf{h}_{F,l}+\sum_{l \in \mathbb{L}_N}\mathbf{h}_{N,l}.\label{hfchannelmodel}
\end{equation}

\subsection{Signal Model}
During uplink channel estimation, the user continuously transmits pilot symbols to the BS for $Q$ time slots. We assume that the channel coherent time is longer than the $Q$ time slots, so that channel state information (CSI) remains static during channel estimation. After the analog beamforming, the equivalent received pilot $\mathbf{y}_q \in \mathbb{C}^{N_{RF}}$ of the BS at time slot $q$ is denoted as
\begin{equation}\label{signal}
	\mathbf{y}_q = \mathbf{W}_{q} \mathbf{h}_{H} x_q+\mathbf{W}_{q} \bm{n}_q,
\end{equation}
where $x_q$ is the transmitted pilot signal at time slot $q$ and
$\mathbf{W}_{q} \in \mathbb{C}^{N_{RF} \times {N}}$ is the beamforming matrix set at the BS.
The term $\bm{n}_q \sim \mathcal{CN}(0, \sigma^{2} \mathbf{I}_{N \times 1})$ is zero-mean complex Gaussian additive noise. Because no prior CSI is available in the channel estimation phase, the beamforming is configured with random phase shifts. 

Based on (\ref{signal}), the received pilot signal at the BS over the entire $Q$ time slots can be written as
\begin{equation}
	\mathbf{y} = \mathbf{W} \mathbf{h}_{H} x+\mathbf{W} \mathbf{n},\label{signalmodel}
\end{equation}
where $\mathbf{y} = [\mathbf{y}_1^T, \mathbf{y}_2^T, ..., \mathbf{y}_Q^T]^T$,  
$\mathbf{W} = [\mathbf{W}_{1}^T, \mathbf{W}_{2}^T, ..., \mathbf{W}_{ Q}^T]^T$, 
and $\bm{n}=[\bm{n}_1^T, \bm{n}_2^T, ..., \bm{n}_Q^T]^T$.

The target of channel estimation is to distinguish each path and estimate the $\{g_l, \theta_l, r_l\}$ for each path of the hybrid-field channel based on the received pilot signal $\mathbf{y}$. Due to a large number of antenna elements $N$, it may lead to a poor channel estimation accuracy if we view (\ref{signalmodel}) as a system of linear equations and solve $\mathbf{h}_{H}$ directly from (\ref{signalmodel}) directly. This is because the channel parameters to be estimated outnumber the pilot signals, i.e. $N>QN_{RF}$.

  
\section{Characteristics of Hybrid-field Channel} \label{sec:transform}
Because of the limited number of scatterers in the millimeter-wave communication\cite{hfomp}, we aim to reduce the channel estimation overhead in (\ref{signalmodel}) based on sparse signal recovery techniques.
In existing works such as ~\cite{classicomp, Polar Domain}, sparse channel characteristics are revealed by transforming channels to the angular domain (or polar domain) for the far-field (or near-field) case. By exploiting the channel sparsity, sparse-signal-recovery-based channel estimation algorithms with low pilot overhead are designed\cite{sparsity}. 

However, in the hybrid-field case, if we apply the angular domain (or polar domain) channel transform, a power diffusion effect occurs. This effect indicates that the power gain of a path spreads to other positions and causes multiple fake paths to be detected. The non-orthogonality between the near-field and far-field steering vectors is the reason for such an effect. Thus, in the hybrid-field case, sparse channel representations can no longer be obtained via either the angular-domain transform or the polar-domain transform, which is shown explicitly below. We therefore apply a joint angular-polar domain channel transform, based on which the power diffusion effect is quantified by the power diffusion range. The power diffusion effect is then alleviated.


\subsection{Introduction to Power Diffusion Effect}\label{subsec:pde}
We first describe the angular-domain and the polar-domain channel transform, where the power diffusion effect of the hybrid-field channel is discovered. Next, the description of such an effect is given.
\subsubsection{Angular-domain Transform} 
The far-field channel is a weighted sum of steering vectors $\mathbf{a}(\theta)$ at different propagation directions of the EM waves, as given in (\ref{ffchannelmodel}). Therefore, a matrix $\mathbf{F}_{A}$ can be designed to transform a far-field channel $\mathbf{h}_{F}$ only consisting of far-field paths to the angular domain representation\cite{farfieldtransform}:
\begin{equation}
	\mathbf{h}_{F} = \mathbf{F}_{A} \mathbf{h}_{A,F},\label{angular}
\end{equation}
where $\mathbf{F}_{A} = [\mathbf{a}(\theta_1), \mathbf{a}(\theta_2), ..., \mathbf{a}(\theta_N)]$ and $\mathbf{a}(\theta_n)$, defined in (\ref{ffsv}), denotes that far-field steering vector toward direction $\theta_n$. We set 
\begin{equation}
\theta_n = \arcsin{\frac{2n-1-N}{N}}, n=1, ..., N. \label{angleset}
\end{equation}
$\mathbf{h}_{A,F}$ is the angular-domain representation of an arbitrary far-field channel $\mathbf{h}_{F}$.

\subsubsection{Polar-domain Transform} 
Similarly, a matrix $\mathbf{F}_{P}$ consisting of near-field steering vectors is designed to transform a near-field channel $\mathbf{h}_{N}$ only consisting of near-field paths to its representation in the polar domain, denoted as
\begin{equation}
	\mathbf{h}_{N} = \mathbf{F}_{P} \mathbf{h}_{P,N},\label{polar}
\end{equation} 
where $\mathbf{F}_{P}$ is obtained by sampling both angles and distances in the space and $\mathbf{h}_{P,N}$ is the polar-domain representation of an arbitrary near-field channel $\mathbf{h}_{N}$. Specifically,  
\begin{align}
& \mathbf{F}_{P} = [\mathbf{F}_{P,1}, \mathbf{F}_{P,2}... \mathbf{F}_{P,S}],\\
& \mathbf{F}_{P,s} = [\mathbf{b}(\theta_1, r_{s,1}), \mathbf{b}(\theta_2, r_{s,2})... \mathbf{b}(\theta_N, r_{s,N})], s = 1, ..., S,
\end{align}
where the term $\mathbf{b}(\theta_1, r_{s,1})$, defined in (\ref{nfsv}), is the near-field steering vector toward position $\{\theta_1, r_{s,1}\}$. The design of $\{\theta_n, r_{s,n}\}$ 
can be found in \cite{Polar Domain}. 

\subsubsection{Power Diffusion Effect}
\begin{figure}[t]
\centerline{\includegraphics[width=6cm,height=4.5cm]{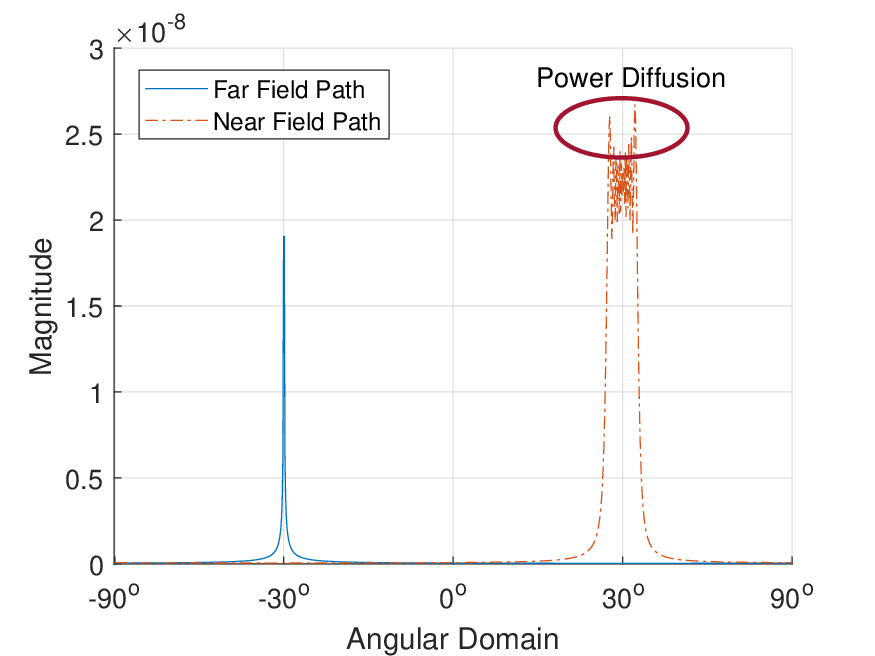}}
\caption{The angular-domain transform result of a hybrid-field channel consisting of a far-field and a near-field steering vector.}
\label{powerdiffusionff}
\end{figure}

\begin{figure}[t]	
\centerline{\includegraphics[width=6cm,height=4.5cm]{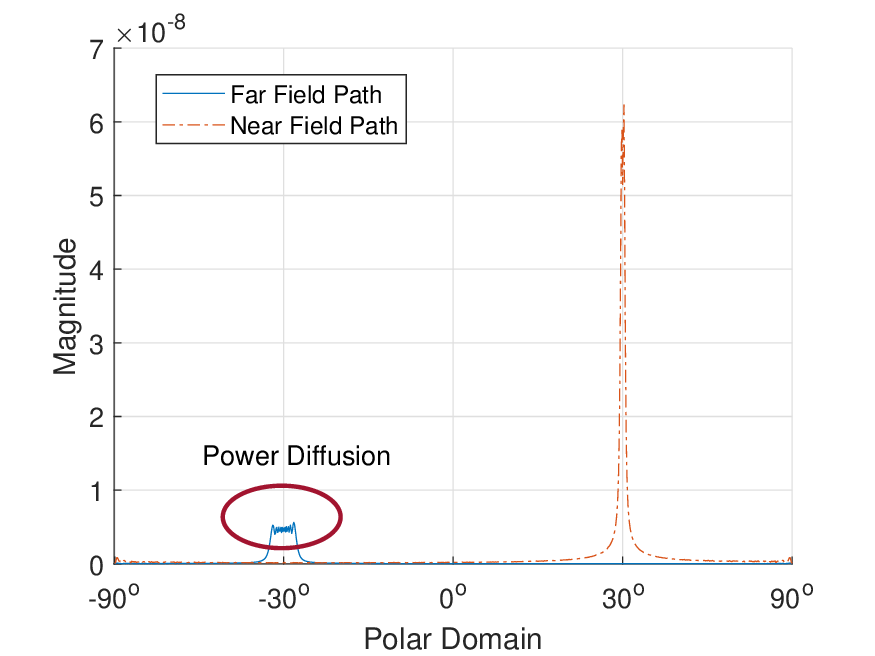}}
\caption{The polar-domain transform result of a hybrid-field channel consisting of a far-field and a near-field steering vector.}
\label{powerdiffusionnf}
\end{figure}

The \emph{Power diffusion effect} refers to the phenomenon that when a path component is transformed from the spatial domain into the angular domain or the polar domain, the power gain corresponding to this path component spreads to other positions and generates fake paths, which are represented by multiple steering vectors. 
To demonstrate the power diffusion effect, in Fig.~\ref{powerdiffusionff} we illustrate the angular-domain representation of a hybrid-field channel $\mathbf{h}_{H}$, which is defined in (\ref{hfchannelmodel}) and consists of a far-field path and a near-field path. The polar-domain representation of this hybrid-field channel is shown in Fig.~\ref{powerdiffusionnf}. 
 They are obtained by
\begin{align}
\mathbf{h}_{A,H} = |\mathbf{F}_{A}^{H} \mathbf{h}_{H}|,\\
\mathbf{h}_{P,H} = |\mathbf{F}_{P}^{H} \mathbf{h}_{H}|.
\end{align}

As shown in Fig.~\ref{powerdiffusionff}, the power gain of the near-field path is not concentrated in one steering vector but spreads over multiple far-field steering vectors in the angular-domain transform matrix $\mathbf{F}_{A}$. Thus, multiple far-field steering vectors should be jointly applied to describe the near-field path. Similarly, in the polar domain, the far-field path should be described by multiple near-field steering vectors, as shown in Fig.~\ref{powerdiffusionnf}. We notice that the power diffusion effect also exists for a near-field path to be transformed to a polar-domain transform matrix consisting of multiple submatrices, i.e., $S \geq 2$, which is not depicted in Fig.~\ref{powerdiffusionnf}. 

\begin{figure}[t]
\centerline{\includegraphics[width=6cm,height=4.5cm]{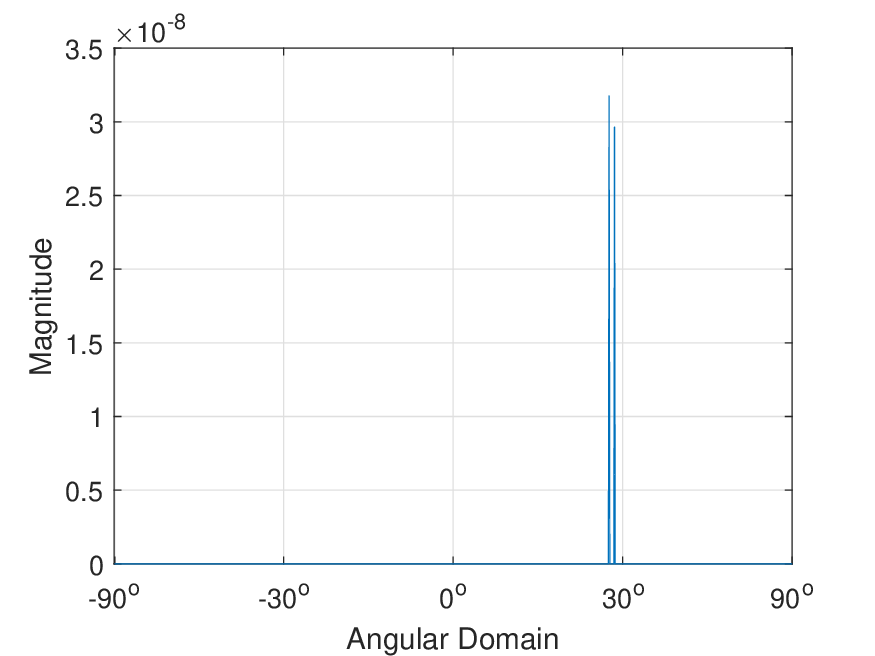}}
\caption{The incorrect path detection result using 2-iteration OMP based on $\mathbf{h}_{A,H}$. The error comes from the power diffusion effect.}
\label{ffomp}
\end{figure}

\begin{figure}[t]	
\centerline{\includegraphics[width=6cm,height=4.5cm]{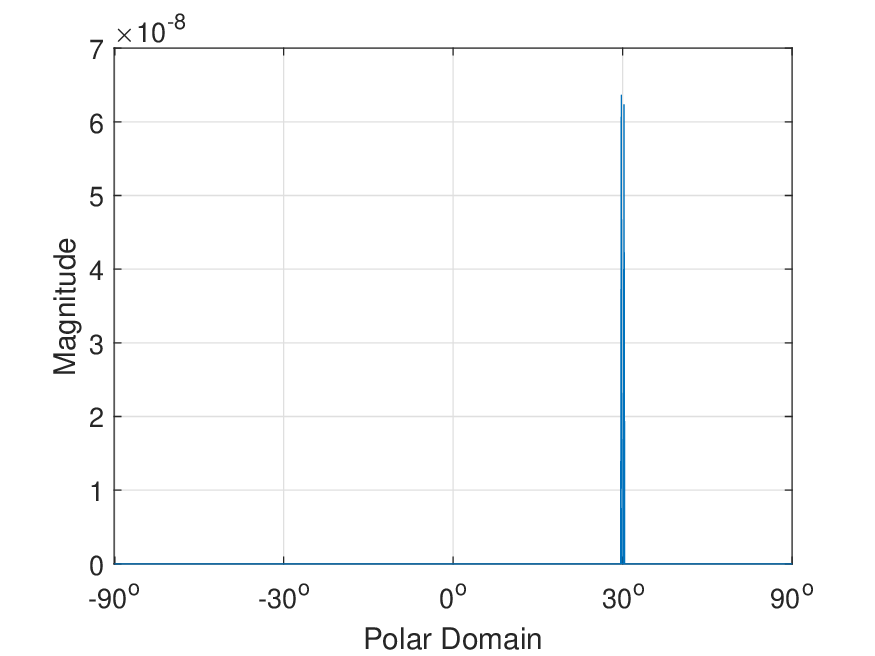}}
\caption{The incorrect path detection result using 2-iteration OMP based on $\mathbf{h}_{P,H}$. The error comes from the power diffusion effect.}
\vspace{-3mm}
\label{nfomp}
\end{figure}

The reason for the power diffusion effect is that two different steering vectors in the angular domain or polar domain can have high coherence. Formally, high coherence can be expressed as, 
\begin{align}
    & \quad \mu_{p,q} = |\mathbf{b}(\theta_p,r_p)^H \mathbf{b}(\theta_q,r_q)| >\alpha, \label{nonn}\\
    & \quad \mu_{r,q} = |\mathbf{a}(\theta_r)^H \mathbf{b}(\theta_q,r_q)| >\alpha,\label{nofn} 
\end{align}
where $\alpha$ is a positive constant, $\mathbf{b}(\theta_p,r_p)$ and $\mathbf{b}(\theta_q,r_q)$ are two different near-field steering vectors
as defined in (\ref{nfsv}), and $\mathbf{a}(\theta_r)$ and is a far-field steering vector
as defined in (\ref{ffsv}). The terms $\mu_{p,q}, \mu_{r,q}$ denote the coherence between two steering vectors. To quantify the power diffusion effect, we define the \emph{power diffusion range} as the range of steering vectors whose coherence with the steering vector representing the path is larger than a pre-defined threshold $\alpha$.


Due to the power diffusion effect in the angular-domain and polar-domain channel representation, the issue of \emph{inaccurate transform-domain path component estimation} arises, which consequently causes errors in spatial-domain channel estimation. 
We illustrate inaccurate transform-domain path component estimation in the angular domain based on conventional OMP channel estimation\cite{samplingbook}.
OMP aims to search for $L$ peaks in the transform-domain channel representation, which represent the $L$ path components of the multipath channel. 
The estimation result of transform-domain channel representation, consisting of the magnitude and corresponding steering vectors for the $L$ peaks, provides sufficient information to restore the CSI.
However, the power diffusion effect causes multiple steering vectors within the power diffusion range, more than $L$ in number, to carry channel information. 
For instance, in Fig.~\ref{powerdiffusionff}, the steering vectors that provide information for the near-field path are within the range of the ellipse. 
Since OMP only identifies $L$ steering vectors, the steering vector representing the far-field path is omitted, resulting in an inaccurate estimation of the transform-domain path component, as shown in  Fig.~\ref{ffomp}.

\subsection{Joint Angular-Polar Channel Transform}

\begin{figure}[t]
	\centerline{\includegraphics[width=9.5cm,height=4.5cm]{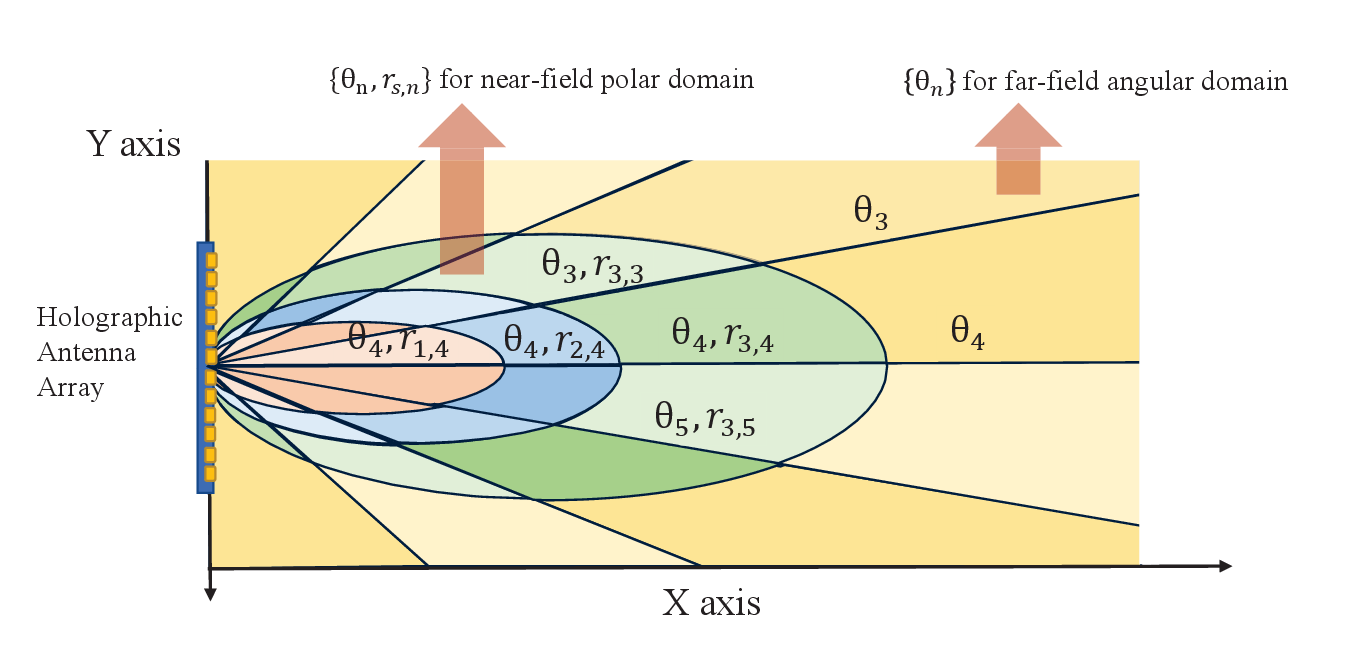}}
	\caption{The illustration of steering vectors in the joint angular-polar domain transform.}
	\label{japd}
        \vspace{-3mm}
\end{figure}

To solve the issue of inaccurate transform-domain path component estimation, we transform the hybrid-field channel to the joint angular-polar domain. Based on this domain, the information of both far-field and near-field paths can be explicitly extracted to eliminate the influence of the power diffusion effect.
The hybrid-field channel representation $\mathbf{h}_{J,H}$ in the joint angular-polar domain is denoted as
\begin{equation}
	\mathbf{h}_{H} = \mathbf{F}_{J}\mathbf{h}_{J,H}.\label{joint}
\end{equation}
 By defining  $\mathbf{F}_{A}, \mathbf{F}_{P,1}, \mathbf{F}_{P,2}... \mathbf{F}_{P,S}$ as the submatrix of $\mathbf{F}_{J}$, the joint angular-polar domain transform matrix $\mathbf{F}_{J}$ can be expressed as~\cite{Polar Domain}
\begin{equation}
	\mathbf{F}_{J} = [\mathbf{F}_{A}, \mathbf{F}_{P,1}, \mathbf{F}_{P,2}... \mathbf{F}_{P,S}].
	\label{transmatrix}
\end{equation}
The far-field and near-field steering vectors contained in $\mathbf{F}_{J}$ are illustrated in Fig.~\ref{japd}.
\begin{figure}[t]
	\centerline{\includegraphics[width=9.5cm,height=5cm]{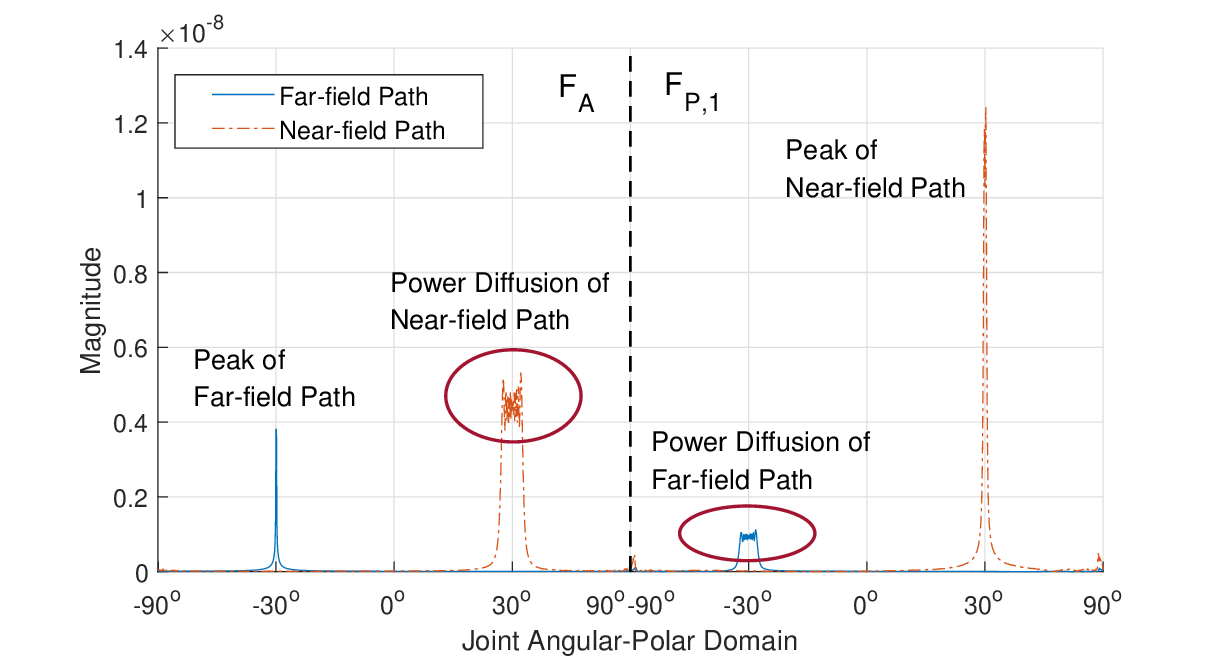}}
	\caption{The channel representation $\mathbf{h}_{J,H}$ in the joint angular-polar domain.}
	\label{powerdiffusion}
\end{figure}

Based on $\mathbf{h}_{J}$, the joint angular-polar domain channel representation of the hybrid-field channel is obtained by
\begin{equation}
\mathbf{h}_{J,H} = |\mathbf{F}_{J}^{H} \mathbf{h}_{H}|.
\end{equation}
An example of a two-path hybrid-field channel representation $\mathbf{h}_{J,H}$ is shown in Fig.~\ref{powerdiffusion}, which is the same channel as shown in Fig.~\ref{powerdiffusionff} and Fig.~\ref{powerdiffusionnf}. Based on Fig.~\ref{powerdiffusion}, we obtain the following two observations.
\begin{observation}\label{ob1}
Given a properly designed joint angular-polar domain transform matrix $\mathbf{F}_{J}$, for each of the $L$ path components, a peak in the joint angular-polar domain corresponding to this path exists.
\end{observation}
\begin{observation}\label{ob2}
In the joint angular-polar domain channel representation, for each peak, the magnitude of its peak is larger than the magnitude of its power diffusion counterpart. This is because, among all steering vectors in $\mathbf{F}_{J}$, the steering vector representing the peak is most correlated with the steering vector representing the path.
\end{observation}

Based on the aforementioned observations, a novel channel estimation method is developed, involving $L$ iterations. Specifically, in the $l$-th iteration, given Observation~\ref{ob1} and Observation~\ref{ob2}, a steering vector representing the peak can be found in the joint angular-polar domain channel representation, indicating that a path component is detected. 
Thus, the estimation of the direction and distance of this path is obtained, and the power diffusion range of the detected path (shown as red ellipses in Fig.~\ref{powerdiffusion}) can be further determined via calculation, which will be described in Section~\ref{sec:algorithm}.
By identifying and eliminating the interference caused by steering vectors within the power diffusion range of the previously detected paths, the $(l+1)$-th iteration can be performed to detect the corresponding steering vector without being influenced by the power diffusion. 
Consequently, this approach resolves the issue of inaccurate transform-domain path component estimation and provides a high-resolution approximation for the joint angular-polar domain channel representation $\hat{\mathbf{h}}_{J}$, thereby enhancing the channel estimation accuracy. The details of the proposed channel estimation method will be elaborated in Section~\ref{sec:algorithm}.


\section{Hybrid-field Channel Estimation Algorithm}\label{sec:algorithm}
In this section, we first propose a hybrid-field channel estimation algorithm without prior knowledge of the number of near-field and far-field paths. We then analyze the computational complexity of the algorithm and the Cram\'er-Rao Lower Bound (CRLB) of the channel estimation problem in the form of sparse signal recovery.

\subsection{Algorithm Design}\label{subsec:algode}
We design a new hybrid-field power-diffusion-aware OMP channel estimation algorithm (PD-OMP), which considers the aforementioned power diffusion effect to improve estimation accuracy. 
\subsubsection{Initialization}
The PD-OMP is first initialized in Steps 1-3. In Step 1, to explicitly reveal the peaks of all far-field and near-field path components of the hybrid-field multipath channel, we generate the transform matrix $\mathbf{F}_{J}$ for the joint angular-polar domain according to its definition in (\ref{transmatrix}).
In Step 2, to whiten the noise in the received signal, we calculate the pre-whitening matrix $\mathbf{D}$ based on the beamforming matrix. 
To be specific, $\mathbf{D}$ is obtained by decomposing the covariance matrix of noise with Cholesky factorization, which is denoted as,
\begin{equation}
\mathbf{C} = \sigma^2 \mathbf{D}\mathbf{D}^H.
\end{equation}
The covariance matrix of noise is computed as~\cite{wn}
\begin{equation}
\mathbf{C} = \sigma^2  \mathbb{B}(\mathbf{W}_1\mathbf{W}_1^H, \mathbf{W}_2\mathbf{W}_2^H, ..., \mathbf{W}_Q\mathbf{W}_Q^H),
\end{equation}
where $\mathbb{B}(\cdot)$ represents the generation of a block diagonal matrix and $\mathbf{W}_q$ is the beamforming matrix of the $q$-th time slot.

In Step 3, we set the equivalent measurement matrix as 
\begin{equation}
\mathbf{\Phi}=\mathbf{D}^{-1}\mathbf{WF}_{J}, 
\end{equation}
 where $\mathbf{\Phi}$ transforms the pilot signal to the joint angular-polar domain in Step 4. 

\subsubsection{Main Body}
\setlength{\textfloatsep}{0cm}
\vspace{0.1in}
\begin{algorithm}[tp]
	\label{alg:pdomp}
	
	\caption{Hybrid-field Channel Estimation Using PD-OMP} 
	\hspace*{0.02in} {\bf Input:} Received pilot signal $\mathbf{y}$, power diffusion detection threshold $\alpha$, number of sampled distances $S$ of $\mathbf{F}_P$, number of paths $L$, beamforming matrix $\mathbf{W}$. 
	\begin{algorithmic}[1]
		\State \textbf {Step 1}. Generate the hybrid-field transform matrix $\mathbf{F}_{J}$ with $S$ based on (\ref{transmatrix}).
        \State \textbf {Step 2}. Calculate the pre-whitening matrix $\mathbf{D}$.
        \State \textbf {Step 3}. Set the equivalent measurement matrix as $\mathbf{\Phi}=\mathbf{D}^{-1}\mathbf{WF}_{H}$. Initialize the support set $\Gamma = \{ \emptyset \}$ and the residual signal $\mathbf{R} = \mathbf{D}^{-1}\mathbf{y}$.
		\For{$l$ $= 1, 2, ..., L$}
		\State \textbf {Step 4}. Detect the $i^{*}_l$-th steering vector to represent path $l$ based on (\ref{detect}) and obtain the corresponding direction and distance  $\{ \theta_{i^{*}_l}, r_{i^{*}_l}\}$. 

		\State \textbf {Step 5}. Generate the power diffusion range $\Gamma_{l}$ for path $l$ with \textbf{Algorithm 2}.
		\State \textbf {Step 6}. Update the support set $\Gamma$ = $\Gamma \cup \Gamma_{l}$.
        \State \textbf {Step 7}. Estimate the sparse channel representation $\hat{\mathbf{h}}_{J,H} = \{\mathbf{\Phi(:,\Gamma)}\}^{\dagger}{\mathbf{y}}.$
		\State \textbf {Step 8}. Update the residual signal as (\ref{signalupdate}). 
		\EndFor
		\State \textbf {Step 9}. Compute the estimated CSI as (\ref{channelrecovery}).
	\end{algorithmic}
	\hspace*{0.02in} {\bf Output:} 
	The estimated CSI $\hat{\mathbf{h}_{H}}$.
\end{algorithm}

The key procedures of PD-OMP are performed iteratively as follows. 
\begin{enumerate}
    \item Path detection (Step 4): Identify the steering vector that exhibits the highest correlation with the residual
pilot signal $\mathbf{R}$ for detecting a path. $\mathbf{R}$ is obtained by subtracting the power gain of detected paths from the received pilot signal.
\item Power diffusion range identification (Step 5): Generate the power diffusion range of the newly detected path.
\item Residual signal update (Steps 6-8): Eliminate the power gain of the newly detected path in the residual signal.
\end{enumerate}

After the initialization stage, $L$ iterations are performed to find the steering vectors corresponding to the $L$ path components from the user to the antenna array. 
Specifically, in Step 4, we first transform the residual signal to the joint angular-polar domain and then detect a new steering vector as 
\begin{equation}
	i^{*}_l = \mathop{\arg\max}_{i}|\mathbf{\Phi}(:,i)^H \mathbf{R}|^2, \label{detect}
\end{equation}
which indicates that the residual signal $\mathbf{R}$ has the strongest correlation with the $i_l^{*}$-th steering vector in $\mathbf{F}_{J}$. In this way, a path is detected, and we would like to point out that the direction and distance $\{ \theta_{i_l^{*}}, r_{i_l^{*}}\}$ associated with the $i_l^{*}$-th steering vector is the estimation result for the propagation direction and distance of the newly detected path.

Given $\{ \theta_{i_l^{*}}, r_{i_l^{*}}\}$ as well as the power gain of the newly detected path, the power diffusion range $\Gamma_{l}$ of this path is generated using \textbf{Algorithm 2} in Step 5, which will be presented in Section~\ref{subsubsec: ipdr}.
The overall support set $\Gamma$ is then updated with the union of $\Gamma_{l}$ in Step 6. The channel representation $\hat{\mathbf{h}}_{J,H}$ is estimated with the least square method in Step 7. The residual signal is updated by removing the projection of the detected paths in the received pilot signal in Step 8 as 
\begin{equation}
\mathbf{R} = {\mathbf{y}} - \mathbf{\Phi}(:,\Gamma)\hat{\mathbf{h}}_{J,H}.\label{signalupdate}
\end{equation}
Finally, in Step 9, the iteration is terminated and the hybrid-field channel is recovered as 
\begin{equation}
\hat{\mathbf{h}_{H}} = \mathbf{F}_{J}\hat{\mathbf{h}}_{J,H}.\label{channelrecovery}
\end{equation}
The proposed PD-OMP channel estimation algorithm is summarized in \textbf{Algorithm 1}.
\subsubsection{Identifying the Power Diffusion Range}\label{subsubsec: ipdr}
\begin{figure}[t]
	\centerline{\includegraphics[width=9cm,height=4.5cm]{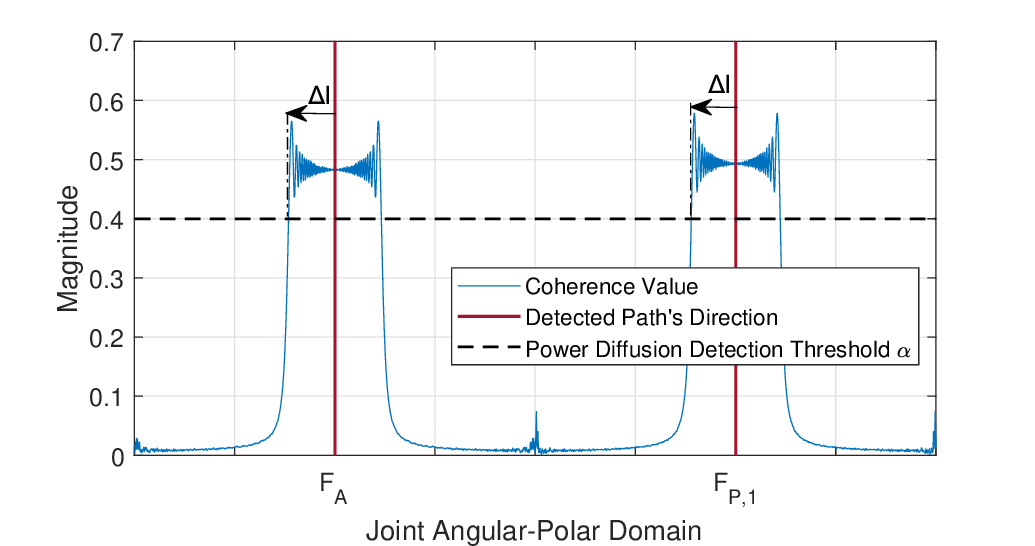}}
        \vspace{-3mm}
	\caption{The illustration of power diffusion range generation based on Algorithm 2.}
         \vspace{5mm}
	\label{fig:varidirect}
\end{figure}

\begin{algorithm}[t]
	\label{alg:pdfg}
	
	\caption{Power Diffusion Range Generation for Path $l$} 
	\vspace*{0.02in}\hspace*{0.02in} {\bf Input:} Hybrid-field transform matrix $\mathbf{F}_{J}$, the normalized magnitude and steering vector index of the $l$-th path $\{ \bar{m}_{l}, i_l^{*}\}$, power diffusion detection threshold $\alpha$. 
	\begin{algorithmic}[1]
		\State \textbf {Initialize} the power diffusion range $\Gamma_l = \{ \emptyset \}$.
		\For{$s$ $= 0, 1, ..., S$}
			\For{$\Delta i$ $=  0, 1, ..., N/2$}
			\State $i_1 = sN + {\rm mod}_N(i_l^{*}) + \Delta i.$
			\State $i_2 = sN + {\rm mod}_N(i_l^{*}) - \Delta i.$
			\State \textbf{Check} the validity of $i_1$ and $i_2$. 
			\State \textbf{Compute} the coherence $\mu_1$ between steering vector $i_1$ and $i_l^{*}$ and the coherence $\mu_2$ between steering vector $i_2$ and $i_l^{*}$. 
			\If{$ \mu_1 \geq \alpha/\bar{m}_{l}$}
				$\Gamma_l$ = $\Gamma_l \cup i_1.$
			\EndIf
			\If{$ \mu_2 \geq \alpha/\bar{m}_{l}$}
				$\Gamma_l$ = $\Gamma_l \cup i_2.$
			\EndIf
			\If{$ \mu_1 < \alpha/\bar{m}_{l}$ and $ \mu_2 < \alpha/\bar{m}_{l}$}
			Break.
			\EndIf
			\EndFor
		\EndFor
	\end{algorithmic}
	\hspace*{0.02in} {\bf Output:} 
	Power diffusion range $\Gamma_l$ for the path $l$.
\end{algorithm}
In Step 5, the power diffusion range of the detected path is generated using \textbf{Algorithm 2}. Specifically, 
the coherence between two steering vectors that are close to each other in direction is generally larger than the coherence between two steering vectors that differ greatly from each other in direction. Hence, to reduce the computational complexity of calculating the coherence, we start by computing the coherence between the $i_l^{*}$-th steering vector and the steering vector which is in each submatrix of $\mathbf{F}_{J}$ and has the same direction as the $i_l^{*}$-th steering vector, i.e., the permitted variation in direction $\Delta i = 0$. Each submatrix of $\mathbf{F}_{J}$ represents a way of sampling the space with a series of steering vectors. The criterion for checking whether the $i$-th steering vector is in the power diffusion range is given as 
\begin{equation}
\mu_{i_l^{*},i} \geq \frac{\alpha}{\bar{m}_{l}},\label{criterion}
\end{equation}
where $\alpha$ is the power diffusion detection threshold and $0 \leq \alpha \leq 1$. $\mu_{i_l^{*},i}$ is the coherence between the $i$-th and $i_l^{*}$-th steering vectors. $\bar{m}_{l}$ is the normalized magnitude of the $l$-th detected path and is defined as 
\begin{equation}
\bar{m}_{l} = \frac{m_{l}}{\max_{l}m_l} = \frac{m_{l}}{m_1} = \frac{\max_{i}|\mathbf{\Phi}(:,i)^H \mathbf{R}|}{m_1},
\end{equation}
where $m_1$ is the magnitude of the first detected path\footnote{Since PD-OMP detects the steering vector with the largest magnitude in each iteration, ${\max_{l}m_l} = m_1$.}. The magnitude of each path is normalized and is introduced into (\ref{criterion}) because the magnitude variation of each path should be considered to limit the power diffusion range of weak paths.
If a steering vector satisfies (\ref{criterion}), this steering vector carries information about the detected path and is therefore added to the power diffusion range. 
To generate the complete power diffusion range, the permitted variation in direction, represented as $\Delta i$, expands in each submatrix until it violates the criterion (\ref{criterion}), as shown in Fig.~\ref{fig:varidirect}.
After we iterate through all the submatrix of $\mathbf{F}_{J}$, the power diffusion range for the detected path is generated with a smaller computational complexity than calculating the coherence between the $i_l^{*}$-th steering vector and each steering vector in $\mathbf{F}_{J}$.

Based on (\ref{criterion}), a smaller $\alpha$ indicates that a wider range of power diffusion is considered. On the one hand, by applying a large $\alpha$, the steering vector representing a path with weak power gain is likely overlooked. On the other hand, small $\alpha$ gives rise to a large power diffusion range, which introduces noise into the estimation result. Therefore, a trade-off of channel estimation accuracy exists for $\alpha$. 
The relationship between $\alpha$ and the size of the power diffusion range will be analyzed in Section \ref{subsec:PDRA} and the effect of $\alpha$ on the performance of PD-OMP will be investigated through simulation in Section \ref{subsec:alphasim}.

\subsection{Computational Complexity Analysis}\label{subsec:cca}
The computational complexity of \textbf{Algorithm 1} mainly comes from
Step 5 and Step 7. The size of matrices or vectors used in PD-OMP is given as $\mathbf{D} \in \mathbb{C}^{N_{RF} \times N_{RF}}$, $\mathbf{\Phi} \in \mathbb{C}^{QN_{RF} \times N(S+1)}$, $\mathbf{R} \in \mathbb{C}^{QN_{RF}}$, $\mathbf{y} \in \mathbb{C}^{QN_{RF}}$ and $\hat{\mathbf{h}}_{J} \in \mathbb{C}^{N(S+1)}$. 

For Step 3, because the matrix inversion can be performed offline, we focus on the complexity of matrix production, which is $\mathcal{O}((QN_{RF})^2)$. The complexity for Step 4, including the matrix product and the maximizing operation, is $\mathcal{O}(NL(S+1)(QN_{RF}+1))$. For Step 5, the computational complexity for coherence calculation is $\mathcal{O}(N)$. Hence, the process of power diffusion range generation requires a complexity of $\mathcal{O}(N(S+1){\rm card}(\Gamma))$, where ${\rm card}(\cdot)$ denotes the number of elements for a given set. Step 6 has a complexity of $\mathcal{O}(L)$. Step 7, 8, 9 have the computational complexities of $\mathcal{O}(QN_{RF}({\rm card}(\Gamma))^2)$, $\mathcal{O}(QN_{RF}{\rm card}(\Gamma))$ and $\mathcal{O}(N{\rm card}(\Gamma))$, respectively. Considering the large dimension of the antenna array and the limited size of the support set, we have $N > QN_{RF}$ and $NLS>({\rm card}(\Gamma))^2$. The computational complexity of PD-OMP is $\mathcal{O}(N(S+1)(LQN_{RF}+{\rm card}(\Gamma)))$, which is linear with the number of antenna elements $N$.

\subsection{Power Diffusion Range Analysis}\label{subsec:PDRA}
Since the size of the power diffusion range, i.e., ${\rm card}(\Gamma)$, directly affects the computational complexity of PD-OMP, we analyze the influence of power diffusion detection threshold $\alpha$ on ${\rm card}(\Gamma)$. 
A closed-form relationship between ${\rm card}(\Gamma)$ and $\alpha$ is intractable because the criterion (\ref{criterion}) needs to be checked for each steering vector in the transform matrix $\mathbf{F}_{J}$ to generate $\Gamma$.
Since the total power diffusion range can be approximated by the sum of the power diffusion range of each path component in each submatrix, i.e., 
\begin{equation}
    {\rm card}(\Gamma) \approx \sum_{l=1}^{L}\sum_{s=1}^{S}{\rm card}(\Gamma_{l,s}),
\end{equation}
we can obtain an approximation for ${\rm card}(\Gamma_{l,s})$ as follows. 

\begin{lemma}\label{lemma_powersum1}
If a path component, which is represented by a steering vector $\mathbf{b}(\theta_p, r_p)$, is transformed based on a submatrix $\mathbf{F}_{J,s} = \left[ \mathbf{b}(\theta_1, r_{s,1}), \mathbf{b}(\theta_2, r_{s,2}), ..., \mathbf{b}(\theta_N, r_{s, N})\right]$, the sum of squares of all transform result components $\left|\mathbf{b}(\theta_n, r_{s,n})^H\mathbf{b}(\theta_p, r_p) \right|, n=1,2,..., N$ approximates to $1$. In other words, the transform-domain representation of this path component satisfies
\begin{equation}
\sum_{n=1}^{N}\left|\mathbf{b}(\theta_n, r_{s,n})^H\mathbf{b}(\theta_p, r_p) \right|^2 \approx 1.
\end{equation}
\end{lemma}
\begin{proof}
    See Appendix \ref{app:lemma_powersum1}.
\end{proof}

\begin{proposition}\label{prop:gammal}
The size of a power diffusion range for a path $l$, i.e., ${\rm card}(\Gamma_l)$, can be approximated as a piece-wise function of $\alpha$:
\begin{equation}
{\rm card}(\Gamma_l) \approx \sum_{s=-S_0+1}^{S-S_0+1}\frac{\epsilon(\bar{m_l}\mu_{l}(s)-\alpha)}{(\mu_{l}(s))^2},
\end{equation}
where $\epsilon(x) = 1$ if $x \geq 0$ and $\epsilon(x) = 0$ if $x < 0$. Here $S_0$ is the index of the submatrix containing the detected $i_l^*$-th steering vector obtained by (\ref{detect}), and $\mu_{l}(s)$ is the coherence between two steering vectors that are in the same direction but are in the submatrices of $\mathbf{F}_{J, S_0}$ and $\mathbf{F}_{J, S_0+s}$, respectively:  
\begin{equation}
\mu_{l}(s) = \frac{1}{N} \left| \sum_{(1-N)/2}^{(N-1)/2} e^{j\frac{2\pi}{\lambda}n^2d^2s\rho} \right|,
\end{equation}
where $\rho$ is a parameter for polar-domain transform matrix design.
\end{proposition}
\begin{proof}

\begin{figure}[t]
	\centerline{\includegraphics[width=8cm,height=6cm]{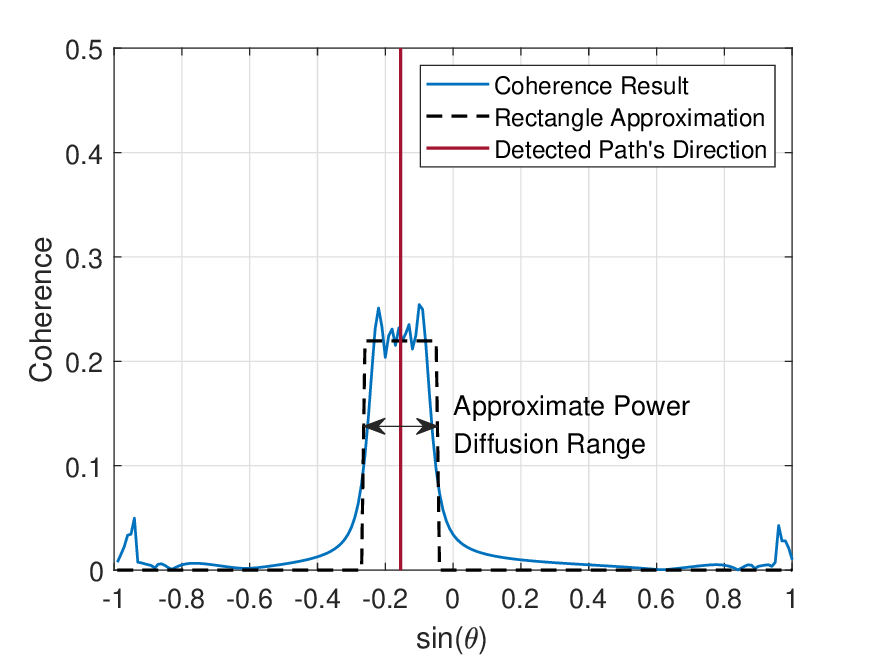}}
	\caption{The illustration of the rectangle approximation of power diffusion range.}
        \vspace{5mm}
	\label{rectapprox}
\end{figure}

 According to the simulation result shown in Fig.~\ref{rectapprox}, the power gain of a detected path concentrates on a few steering vectors around the direction of the detected path. Hence, we apply a rectangle to approximate the power diffusion effect in the $(s-S_0)$-th submatrix such that ${\rm card}(\Gamma_{l,s})$ is approximately equal to the width of the rectangle $\Delta D$. We set the steering vector which has the same direction as the detected  $i_l^*$-th steering vector to be the midpoint of the rectangle. The height of the rectangle is therefore set as $\mu_l(s)$. According to \textbf{Lemma} \ref{lemma_powersum1}, we have
\begin{align}
& {\rm card}(\Gamma_{l,s}) \approx \Delta D  \approx \nonumber \\ 
& \frac{1}{(\mu_l(s))^2} \sum_{n=1}^{N}\left|\mathbf{b}(\theta_n, r_{s,n})^H\mathbf{b}(\theta_p, r_p) \right|^2 \approx \frac{1}{(\mu_l(s))^2}.
\end{align}

Since the normalized magnitude of each path is considered in the criterion of power diffusion range (\ref{criterion}), condition $\bar{m_l}\mu_{l}(s) \geq \alpha$ is introduced to limit the power diffusion range, which concludes the proof.
\end{proof}
Based on \textbf{Proposition} ~\ref{prop:gammal}, the relationship between the support set $\Gamma$ and the power diffusion detection threshold $\alpha$ is given as
\begin{equation}
    {\rm card}(\Gamma) \approx \sum_{l=1}^{L}\sum_{s=-S_0+1}^{S-S_0+1}\frac{\epsilon(\bar{m_l}\mu_{l}(s)-\alpha)}{(\mu_{l}(s))^2}.
\end{equation}
\subsection{Cram\'er-Rao Lower Bound Analysis}
The CRLB bound serves as a theoretical lower bound of MSE to evaluate the performance of channel estimation algorithms. We first derive the CRLB for the estimation of sparse channel representation $\hat{\mathbf{h}}_{J,H}$. Then we obtain the CRLB for the spatial-domain hybrid-field channel $\hat{\mathbf{h}}_{H}$ based on $\hat{\mathbf{h}}_{H} = \mathbf{F}_{J}\hat{\mathbf{h}}_{J,H}$.

\begin{lemma}
    The CRLB for the estimation of sparse channel representation $\hat{\mathbf{h}}_{J}$ is given as \cite{sparseCRLB}
    \begin{equation}\label{lemmasparseCRLB}
    \mathbb{E}\left\{ \Vert \hat{\mathbf{h}}_{J,H} - \mathbf{h}_{J,H}\Vert_2^2 \right\} = \sigma^2{\rm Tr} \left ( \left ( \mathbf{\Phi}_{\mathbf{h}_{J,H}}^H \mathbf{\Phi}_{\mathbf{h}_{J,H}}\right )^{-1} \right ),
    \end{equation}
    where $\mathbf{\Phi}_{\mathbf{h}_{J,H}} \in \mathcal{C}^{QN_{RF}\times {\rm card}(\Gamma)}$ is a matrix composed of columns of $\mathbf{\Phi}$ indexed by the indices of true support set of $\mathbf{h}_{J,H}$.  
\end{lemma}
Since ${\rm Rank}(\mathbf{\Phi}_{\mathbf{h}_{J,H}}^H \mathbf{\Phi}_{\mathbf{h}_{J,H}})={\rm card}(\Gamma)$, (\ref{lemmasparseCRLB}) can be further written as
\begin{equation}
{\rm Tr} \left ( \left ( \mathbf{\Phi}_{\mathbf{h}_{J,H}}^H \mathbf{\Phi}_{\mathbf{h}_{J,H}}\right )^{-1} \right )=\sum_{i=1}^{{\rm card}(\Gamma)} \lambda_i^{-1},
\end{equation}
where $\lambda_1, \lambda_2, ... \lambda_{{\rm card}(\Gamma)}$ are the eigenvalues of $\mathbf{\Phi}_{\mathbf{h}_{J,H}}^H \mathbf{\Phi}_{\mathbf{h}_{J,H}}$.

The coherence of $\mathbf{\Phi}$ is defined as
\begin{equation}
\mu_{\mathbf{\Phi}} = \max_{i\neq j} \vert\bm{\phi}_i^H\bm{\phi}_j\vert,
\end{equation}
where $\bm{\phi}_i$ and $\bm{\phi}_j$ are the $i$-th and $j$-th column of $\mathbf{\Phi}$, respectively, and have the following forms:
\begin{align}
&\bm{\phi}_i = \mathbf{D}^{-1}\mathbf{W}\mathbf{f}_{i},\label{phii}\\
&\bm{\phi}_j = \mathbf{D}^{-1}\mathbf{W}\mathbf{f}_{j},\label{phij}
\end{align}
where $\mathbf{f}_{i}$ and $\mathbf{f}_{j}$ are the $i$-th and $j$-th column of $\mathbf{F}_{J}$, respectively.
According to the Gershgorin Disc Theorem \cite{matcomp} as well as the fact that $\mathbf{\Phi}_{\mathbf{h}_{J,H}}^H \mathbf{\Phi}_{\mathbf{h}_{J,H}}$ is a positive semidefinite matrix, the eigenvalues of $\mathbf{\Phi}_{\mathbf{h}_{J,H}}^H \mathbf{\Phi}_{\mathbf{h}_{J,H}}$ are real and lie in the range of $\left [  \max\{ 1-{\rm card}(\Gamma)\mu_{\mathbf{\Phi}},0\}, 1+{\rm card}(\Gamma)\mu_{\mathbf{\Phi}}\right ]$. Therefore, we have
\begin{equation}
\mathbb{E}\left\{ \Vert \hat{\mathbf{h}_{J,H}} - \mathbf{h}_{J,H}\Vert_2^2 \right\} \geq \frac{\sigma^2 {\rm card}(\Gamma)}{1+{\rm card}(\Gamma)\mu_{\mathbf{\Phi}}}.
\end{equation}

Since $\bm{\phi}_i$ and $\bm{\phi}_j$ is obtained by the randomly generated beamforming matrix $\mathbf{W}$, as shown in (\ref{phii}) and (\ref{phij}), $\mu_{\mathbf{\Phi}}$ is a random variable with respect to $\mathbf{W}$.
Here we derive an upper bound on the expectation of $\mu_{\mathbf{\Phi}}$. 
\begin{lemma}\label{lemmamubound}
The upper bound for the expectation of $\mu_{\mathbf{\Phi}}$ with respect to $\mathbf{W}$ is given as,
\begin{equation}
    \mathbb{E}\left\{ \mu_{\mathbf{\Phi}} \right\} < \frac{QN_{RF}}{N}.
    \end{equation}
\end{lemma}
\begin{proof}
See Appendix~\ref{app:lemma_mubound}.
\end{proof}

Based on \textbf{Lemma}.~\ref{lemmamubound}, we have
\begin{align}
&\mathbb{E}\left\{ \Vert \hat{\mathbf{h}}_{J,H} - \mathbf{h}_{J,H}\Vert_2^2 \right\} \geq \frac{\sigma^2 {\rm card}(\Gamma)}{1+{\rm card}(\Gamma)\mu_{\mathbf{\Phi}}} \nonumber \\
&\approx \frac{\sigma^2 {\rm card}(\Gamma)}{1+{\rm card}(\Gamma)\mathbb{E}\left\{\mu_{\mathbf{\Phi}}\right\}} > \frac{\sigma^2 {\rm card}(\Gamma)}{1+\frac{{\rm card}(\Gamma)QN_{RF}}{N}}.
\end{align}

Given $\mathbf{h}_{H} = \mathbf{F}_{J}\mathbf{h}_{J,H}$, we have the following proposition.
\begin{proposition} \label{prop:crlb}
The CRLB for the estimated hybrid-field multipath channel $\hat{\mathbf{h}}$ is given as
\begin{equation}
\mathbb{E} \left\{ \Vert \hat{\mathbf{h}}_{H} - \mathbf{h}_{H} \Vert_2^2\right\} \geq \frac{\sigma^2 {\rm card}(\Gamma)}{1+\frac{{\rm card}(\Gamma)QN_{RF}}{N}}(\sigma_{\min}(\mathbf{F}_{J}))^2,
\end{equation}
where $\sigma_{\min}(\mathbf{F}_{J})$ denotes the smallest singular value of $\mathbf{F}_{J}$.
\end{proposition}
\begin{proof}
See Appendix~\ref{app:crlb}. 
\end{proof}

Based on the above formula for CRLB, it is observed that the increase in the number of RF chains and pilot length can reduce the CRLB. This is reasonable as additional RF chains and pilot signals can provide more information on the hybrid-field channel, thus improving the estimation accuracy. Besides, the increase in antenna numbers, which indicates that more parameters have to be estimated, increases the CRLB. The increase in power diffusion range size ${\rm card}(\Gamma)$ also increases the CRLB.

\section{Simulation Results}\label{sec:simulation_results}
In this section, we evaluate the performance of our proposed channel estimation algorithm PD-OMP in terms of the normalized mean square error (NMSE). The NMSE is defined as 
\begin{equation}
{\rm NMSE} = \mathbb{E}\left\{\frac{\Vert \mathbf{\hat{h}}_{H} - \mathbf{h}_{H}\Vert_2^2}{\Vert \mathbf{h}_{H}\Vert_2^2}\right\},
\end{equation}
which is the expectation of the square of the relative estimation error. Besides, the influence of power diffusion detection threshold $\alpha$ on the performance of PD-OMP is investigated.

\subsection{Parameter Setting}

In the simulation, we consider the case where an antenna array at the working frequency of $30$ GHz is equipped with 200 antenna elements.
The number of RF chains is set as $N_{RF} = 10$. The number of paths in the multipath channel is set as $L = 7$. The distance and angle of the user and scatterers to the origin satisfy the uniform distribution and are within the range of $(30m, 300m)$ and $(-60^{\circ}, 60^{\circ})$, respectively. 
The joint factor of channel fading and scattering $g_l$ of NLoS path satisfies circularly symmetric complex Gaussian distribution of $\mathcal{CN}(0,1)$. For the LoS path, $g_l = 1$.
Each element of the beamforming matrix $\mathbf{W}$ is randomly chosen from $\{\frac{1}{\sqrt{N}},-\frac{1}{\sqrt{N}}\} $ with equal probability. The number of submatrices for $\mathbf{F}_P$ is $5$.

In the simulation, we calculate the normalized CRLB of compressed-sensing-based hybrid-field channel estimation, which is denoted as
\begin{equation}
    {\rm CRLB} = \frac{\sigma^2 {\rm card}(\Gamma)}{(1+\frac{{\rm card}(\Gamma)QN_{RF}}{N})\Vert\mathbf{h}_{H}\Vert_2^2}.
\end{equation}
To evaluate the performance of PD-OMP, we also compare it with the basic MMSE algorithm and five OMP-based channel estimation algorithms, i.e.,
\begin{enumerate}    
    \item \emph{MMSE} \cite{MMSE}: A basic estimation method that applies the second-order statistics of the CSI to minimize the mean square error of channel estimation.
    \item \emph{NPD-OMP}: Estimate the far-field and near-field path components simultaneously with joint angular-polar domain transform.
    \item\emph{Near-OMP} \cite{Polar Domain}: Only apply the near-field polar-domain transform (\ref{polar}).
    \item\emph{Far-OMP} \cite{classicomp}: Only apply the far-field angular-domain transform (\ref{angular}).
    \item \emph{HF-OMP} \cite{hfomp}: Require the numbers of near-field and far-field paths as prior information and estimate the far-field and near-field path components separately.  
    \item \emph{SD-OMP} \cite{sdomp}: Require the numbers of near-field and far-field paths as prior information and estimate the far-field and near-field path components separately. When estimating far-field path components, the support set $\Gamma$ also includes far-field steering vectors whose directions are close to the detected far-field steering vector's direction. 
\end{enumerate}
It should be noted that none of the comparing algorithms consider the power diffusion effect.

\begin{table*}[t]
\centering
\caption{Mean and Standard Deviation of NMSE with respect to $\gamma$ for different algorithms.}
\begin{tabular}{|c|c|c|c|c|c|c|}
\hline
NMSE (dB) &\textbf{PD-OMP (Proposed)}  & Near-OMP [11]& Far-OMP [18] & HF-OMP [13]&  MMSE [29]\\
\hline
Mean & \textbf{-6.792 (\checkmark)}		&-3.973	&-5.899	&-5.341	&-2.428\\ 
\hline
Standard Deviation & 0.118 	&0.448	&0.537	&0.683 & \textbf{0.020 (\checkmark)}\\
\hline
Range & \textbf{[-6.964, -6.646] (\checkmark)} & [-4.862, -3.412] & [-6.800, -5.081] & [-6.800, -4.606] & [-2.395, -2.462] \\
\hline
\end{tabular}
\label{tab:stc}
\end{table*}
\subsection{Influence of SNR on Algorithm Performance}
\begin{figure}[t]
	\centerline{\includegraphics[width=8cm,height=6cm]{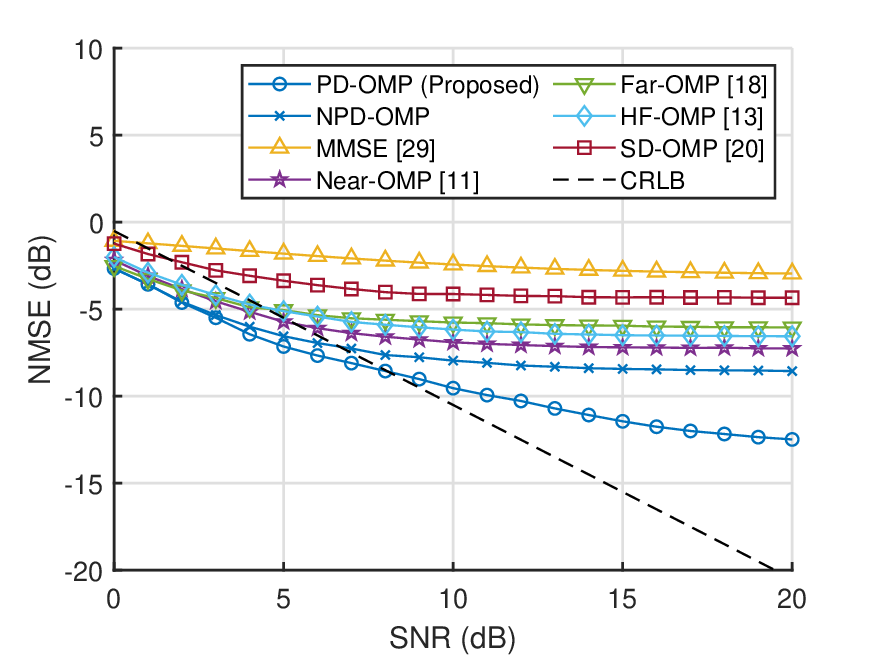}}
        \vspace{-3mm}
	\caption{The NMSE performance versus SNR. The pilot length is set as $Q = 10$. For each SNR, $\alpha$ reaching the minimum NMSE is chosen. }
	\label{pic:nmsevsSNR}
        \vspace{5mm}
\end{figure}
Fig.~\ref{pic:nmsevsSNR} demonstrates the NMSE performance of different estimation algorithms versus SNR.
Compared with HF-OMP or SD-OMP, which requires the numbers of near-field and far-field paths as prior knowledge, PD-OMP can estimate the channel more accurately without the prior knowledge of path distribution, demonstrating the effectiveness of the proposed algorithm. 
This is because PD-OMP applies the joint angular-polar transform matrix, which is capable of capturing both the far-field and near-field features of the channel.
The NMSE of PD-OMP is lower than NPD-OMP, showing the necessity of considering the power diffusion range in the support set.
Besides, PD-OMP enjoys superiority over benchmark algorithms when the SNR is closer to 20 dB.
This is because, at a higher SNR, the power diffusion range can be estimated more accurately, which is then utilized by PD-OMP to compensate for the performance degradation caused by the power diffusion effect. In contrast, none of the existing algorithms consider the power diffusion effect. 

Besides, we calculate the CRLB to evaluate the theoretical performance of PD-OMP. We observe that  CRLB is larger than the NMSE when SNR $\leq 8$ dB, which shows that CRLB is valid for PD-OMP only at a high SNR regime. This is because CRLB gives a theoretical lower bound for unbiased estimators. However, OMP is a biased estimator~\cite{ompbias} and the bias is not negligible at a lower SNR regime.  
\vspace{-4mm}
\subsection{Influence of Pilot Length on Algorithm Performance}
\begin{figure}[t]
	\centerline{\includegraphics[width=8cm,height=6cm]{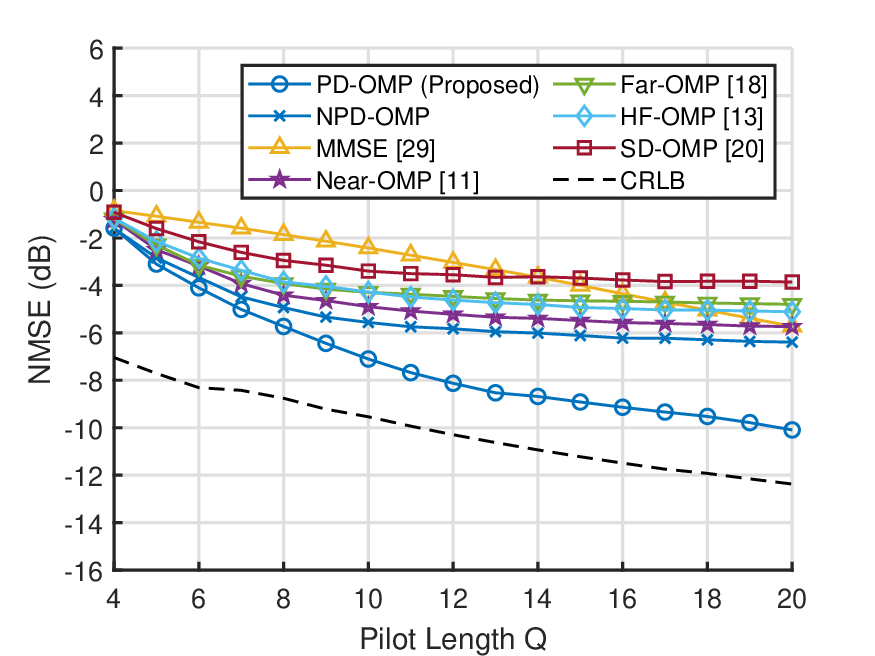}}
 \vspace{-3mm}
	\caption{The NMSE performance versus the pilot length $Q$. SNR is set as $10$ dB. $\alpha$ reaching the minimum NMSE is chosen for each pilot length.}
 \vspace{5mm}
	\label{pic:nmsevsQ}
\end{figure}
Fig.~\ref{pic:nmsevsQ} presents the NMSE performance of different algorithms with the increase of pilot length $Q$. PD-OMP can achieve the lowest NMSE among all benchmark methods for different $Q$. In other words, PD-OMP requires a small pilot length to achieve the same estimation accuracy. For instance, to reach an NMSE of $-5.8$ dB, PD-OMP only requires $Q=8$ while MMSE requires $Q=20$. Besides, the gaps of NMSE between PD-OMP and benchmark methods increase with $Q$. This is because more pilot signals can provide more information on the channel so that for PD-OMP, the power diffusion range can be generated more accurately to help channel estimation.

\vspace{-4mm}
\subsection{Influence of Scatterer Distribution}
The scatterer distribution is defined as the split ratio $\gamma = {\rm card}(\mathbb{L}_N)/L$, which represents the ratio of the near-field path numbers ${\rm card}(\mathbb{L}_N)$ to the total multipath numbers $L$. The mean, standard deviation, maximum, and minimum of NMSE concerning $\gamma$ are applied to evaluate whether algorithms are robust to the variation of the split ratio. The mean and standard deviation of NMSE are calculated by 
\begin{align}
& \overline{\rm NMSE} = \frac{\sum_{\gamma \in \Xi} {\rm NMSE_{\gamma}}}{{\rm Card}(\Xi)},\\
&  \sigma_{\rm NMSE} = \sqrt{\frac{\sum_{\gamma \in \Xi} ({\rm NMSE_{\gamma}} -\overline{{\rm NMSE}})^2} {{\rm Card}(\Xi)}},
\end{align}
where $\Xi = \{0, 0.1,..., 1\}$ and ${\rm NMSE_{\gamma}}$ denotes the NMSE performance obtained by PD-OMP given the split ratio $\gamma$. The simulation parameters are set as $L=10$, SNR = $10$ dB, and $Q$ = 10. 

It is observed from \textbf{Table}~\ref{tab:stc} that PD-OMP achieves the lowest mean estimation error among all benchmark algorithms and the lowest standard deviation of estimation error among all OMP-based benchmark algorithms. It is noted that the MMSE method obtains a lower standard deviation than PD-OMP because MMSE employs the statistical information of the channel to assist channel estimation. Besides, the PD-OMP has the smallest maximum NMSE among all methods. Hence, PD-OMP is robust to the change of split ratio in the hybrid-field wireless communication scenario, demonstrating its stability against the scatterer distribution variation.

\subsection{Advantage of Joint Angular-Polar Domain Channel Transform} \label{subsec:hfvspf}
\begin{figure}[t]
	\centerline{\includegraphics[width=8cm,height=6cm]{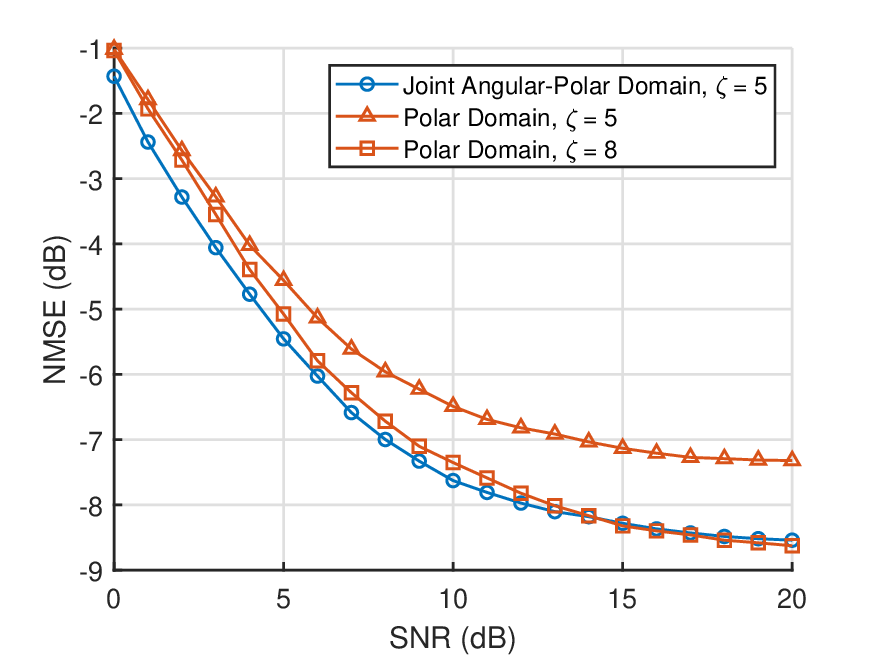}}
 \vspace{-3mm}
	\caption{The NMSE performance of PD-OMP and Near-PD-OMP with different numbers of submatrices $\zeta$ in the transform matrix. We set $Q = 10$ and $\alpha = 0.7$.} 
 \vspace{5mm}
	\label{pic:hfvspf}
\end{figure}
To demonstrate the necessity of applying joint angular-polar domain channel transform instead of the polar-domain channel transform in the hybrid-field channel estimation, we compare the performances of two transforms in Fig.~\ref{pic:hfvspf}. The curves ``Polar Domain $\zeta=5$" and ``Polar Domain $\zeta=8$" are obtained with a channel estimation method that replaces the joint angular-polar domain transform in PD-OMP with the polar-domain transform. We use $\zeta$ to denote the number of submatrices in the transform-domain matrix. A larger $\zeta$ indicates that a larger area is sampled with near-field steering vectors, which induces a higher computational complexity. 

As shown in Fig.~\ref{pic:hfvspf}, when we set $\zeta=5$ for both transform matrices, i.e., the computational complexities for two algorithms based on different transform matrices are the same, the NMSE of the joint angular-polar domain transform is smaller than the polar-domain transform. Compared with the polar-domain channel transform ($\zeta=8$), the joint angular-polar domain channel transform ($\zeta=5$) saves $37.5\%$ computational complexity while achieving a similar NMSE performance.
Hence, the joint angular-polar domain can help reach a balance between the estimation accuracy and the computational complexity.

\subsection{Influence of Power Diffusion Detection Threshold $\alpha$ on Algorithm Performance} \label{subsec:alphasim}

Power diffusion detection threshold $\alpha \in \left(0, 1\right ]$ decides the considered power diffusion range in PD-OMP, which affects the estimation accuracy and the computational complexity of PD-OMP. 

\begin{figure}[t]
	\centerline{\includegraphics[width=8cm,height=6cm]{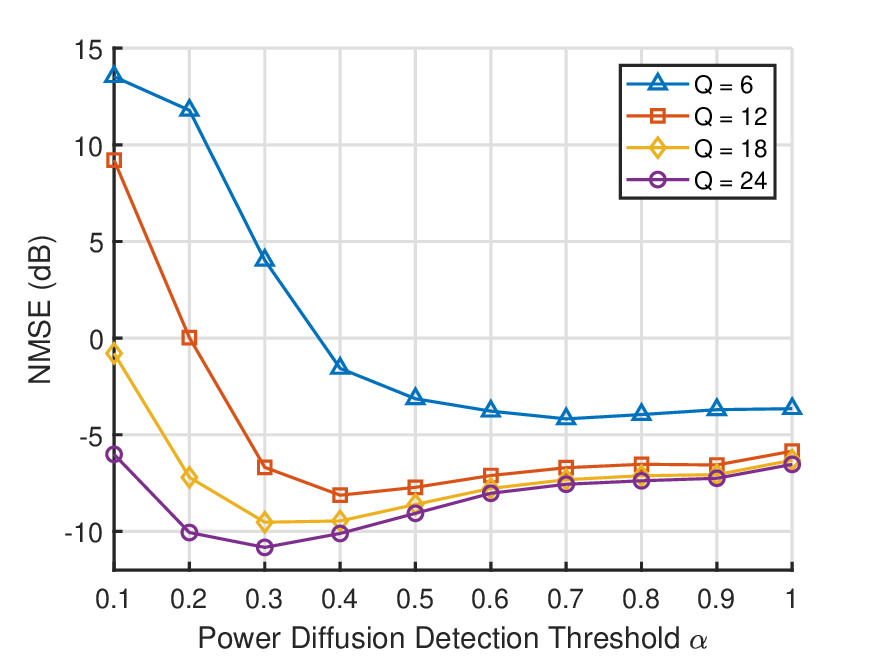}}
 \vspace{-3mm}
	\caption{The NMSE performance of PD-OMP versus different $\alpha$ and different pilot length. SNR is set as $10$ dB.}
	\label{pic:nmsevsalpha2DQ}
  \vspace{3mm}
\end{figure}

In Fig.~\ref{pic:nmsevsalpha2DQ}, we present how the NMSE of PD-OMP changes with the power diffusion detection threshold $\alpha$ when the pilot length $Q$ increases. From Fig.~\ref{pic:nmsevsalpha2DQ}, the NMSE of each $\alpha$ reduces as $Q$ increases. For a higher $Q$, a smaller $\alpha$ reaches the minimum NMSE. 
This is because given more pilot signals, with a smaller $\alpha$, a wider range of considered power diffusion can help eliminate the estimation error more thoroughly. 
Nevertheless, in the case of a small $Q$, a falsely estimated range of power diffusion is likely to be introduced into the support set $\Gamma$ if a small $\alpha$ is adopted. Thus, a high $\alpha$ is desirable with a small $Q$, as the range of power diffusion is limited. 
Fig.~\ref{pic:nmsevsalpha2DQ} also reveals that the pilot length serves as auxiliary information for choosing a proper size of power diffusion range in channel estimation.

\begin{figure}[t]
	\centerline{\includegraphics[width=8cm,height=6cm]{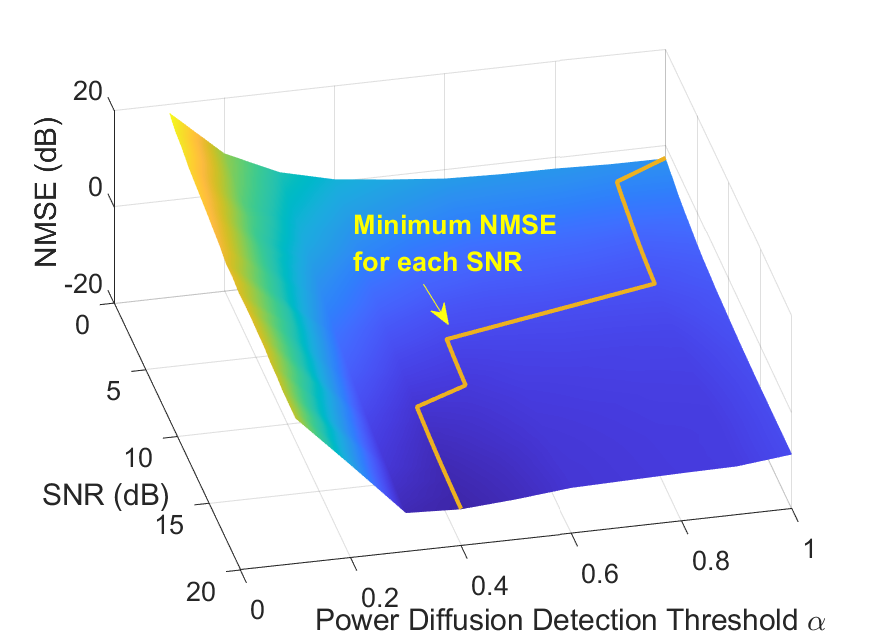}}
 \vspace{-3mm}
	\caption{The NMSE performance of PD-OMP of different $\alpha$ versus SNR. $Q$ is set as $10$.}
	\label{pic:nmsevsalpha3DSNR}
\end{figure}

In Fig.~\ref{pic:nmsevsalpha3DSNR}, how the NMSE of PD-OMP changes with the SNR and the power diffusion detection threshold $\alpha$ is investigated. The NMSE reduces as the SNR increases for different $\alpha$. Besides, according to the yellow curve, with the increase of SNR, PD-OMP with a decreased $\alpha$ can achieve the lowest NMSE.
This is because the inaccurate transform-domain path component estimation is alleviated by applying a smaller $\alpha$ in the case of a high SNR. However, if the SNR is low, a falsely included power diffusion range will introduce additional noise to the estimation result and worsen the NMSE performance. 
Therefore, Fig.~\ref{pic:nmsevsalpha3DSNR} reveals that when the SNR is low, a high $\alpha$ should be selected to limit the range of power diffusion to improve the channel estimation accuracy. 

\begin{figure}[t]
	\centerline{\includegraphics[width=8cm,height=6cm]{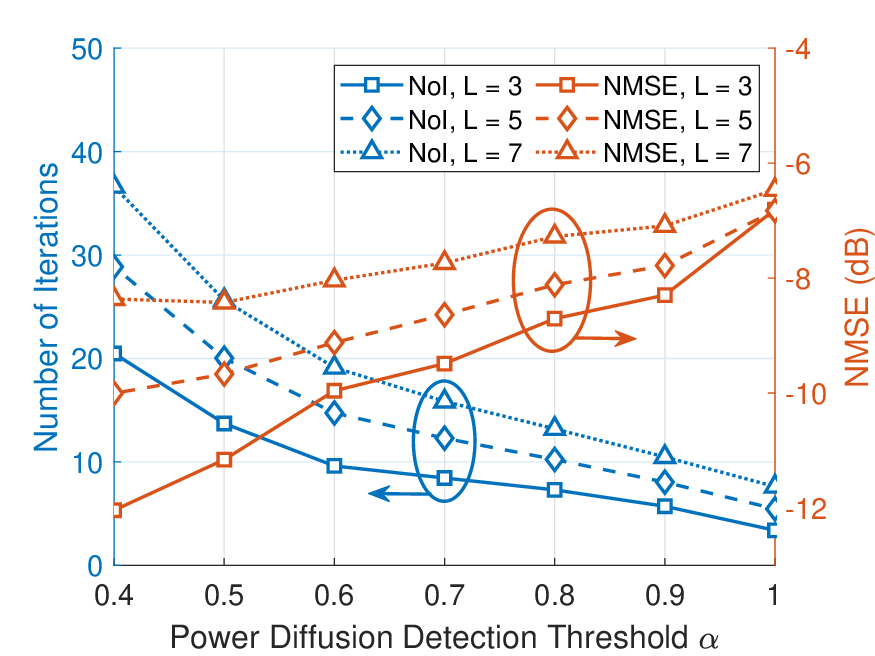}}
 \vspace{-3mm}
	\caption{The number of iterations (NoI) and the NMSE versus the power diffusion detection threshold $\alpha$ for different numbers of multipath $L$ with $Q = 10$ and SNR = $10$ dB. NoI refers to the times Step 4 to Step 10 in \textbf{Algorithm} 2 are performed.}
 \vspace{3mm}
	\label{pic:tradeoff}
\end{figure}

In Fig.~\ref{pic:tradeoff}, the number of iterations and NMSE of PD-OMP concerning $\alpha \in \left\{0.4, ..., 1\right\}$ are investigated. The number of iterations refers to how many times Step 4 to Step 10 in \textbf{Algorithm} 2 are performed, which dominates the computational complexity of PD-OMP. 
First, a trade-off between the computational complexity and the estimation accuracy is revealed. When $\alpha$ rises from 0.4 to 1, the number of iterations reduces as a narrow range of power diffusion is selected, which reduces the calculation of coherence and the matrix inversion in PD-OMP. However, the estimation accuracy also worsens with the increase of $\alpha$.
Second, it is observed that with the increase of paths number, both the number of iterations and NMSE increase. This is because additional paths lead to an increasingly complicated multipath channel and a wider power diffusion range has to be estimated, which gives rise to higher estimation error. 
\section{Conclusion}\label{sec:conclusion}

In this paper, we investigated the hybrid-field multipath channel estimation in holographic MIMO communications. We first identified the issue of inaccurate path component estimation, which led to inaccurate channel estimation if the conventional far-field or near-field channel estimation methods were applied to the hybrid-field case. We revealed that the reason came from the power diffusion effect that the power gain of each path diffused to other positions. In consequence, fake paths are generated in the channel representation in transform domains so that path components were difficult to be separated from each other. To cope with the power diffusion effect, the hybrid-field channel was transformed from the spatial domain to the joint-angular-polar domain. A solution concept of power diffusion range was introduced to quantify the range of diffused power gain so that the power diffusion effect could be identified and eliminated. A novel channel estimation algorithm PD-OMP was then proposed. The computational complexity of PD-OMP and the CRLB of sparse-signal-recovery-based hybrid-field channel estimation were derived. The theoretical analysis showed that the computational complexity of PD-OMP was linear with the number of antenna elements.

Simulation results showed that: 1) By extracting both far-field and near-field path components of the hybrid-field channel, the joint angular-polar domain channel transform reached a balance between the estimation accuracy and the computational complexity compared to the polar-domain channel transform. 2) The proposed PD-OMP outperformed current state-of-the-art hybrid-field channel estimation methods under different SNRs and pilot lengths. Besides, PD-OMP was robust to the variation of scatterer distribution. 3) When the SNR or the pilot length increased, a wider range of power diffusion should be selected in PD-OMP to produce a smaller NMSE. 4) There existed an optimal power diffusion detection threshold to reach a trade-off between the computational complexity and the channel estimation accuracy of PD-OMP.

\vspace{-2mm}
\begin{appendices}
\section{Proof of Lemma \ref{lemma_powersum1}}\label{app:lemma_powersum1}
We omit the propagation direction and transmission distance $(\theta_p, r_p)$ of the steering vector $\mathbf{b}(\theta_p, r_p)$ for simplicity. Hence, we have
\begin{align}
& \sum_{n=1}^{N}\left|\mathbf{b}(\theta_n, r_{s,n})^H\mathbf{b} \right|^2 = (\mathbf{F}_{J,s}\mathbf{b})^H\mathbf{F}_{J,s}\mathbf{b} \nonumber \\
& = {\rm Tr}(\mathbf{F}_{J,s}^H\mathbf{F}_{J,s}\mathbf{b}\mathbf{b}^H),
\end{align}
where $\mathbf{F}_{J,s}$ is the $s$-th submatrix of $\mathbf{F}_{J}$. The term $\mathbf{F}_{J,s}^H\mathbf{F}_{J,s}$ is denoted as
\vspace{-2mm}
\begin{equation}
\mathbf{F}_{J,s}^H\mathbf{F}_{J,s} = \begin{pmatrix}
\chi_{1,1} & \cdots & \chi_{1,N} \\
\vdots & \ddots & \vdots \\
\chi_{N,1} & \cdots & \chi_{N,N}
\end{pmatrix},
\end{equation}
where the $(x,y)$-th element of $\mathbf{F}_{J,s}^H\mathbf{F}_{J,s}$ is expressed as $\chi_{x,y} = \mathbf{b}(\theta_x, r_{s,x})^H\mathbf{b}(\theta_y, r_{s,y})$. The term $\mathbf{b}\mathbf{b}^H$ is denoted as
\vspace{-2mm}
\begin{equation}
\mathbf{b}\mathbf{b}^H = \frac{1}{N}\begin{pmatrix}
e^{-jk(r_{1,p}-r_{1,p})} & \cdots & e^{-jk(r_{1,p}-r_{N,p})} \\
\vdots & \ddots & \vdots \\
e^{-jk(r_{N,p}-r_{1,p})} & \cdots & e^{-jk(r_{N,p}-r_{N,p})}
\end{pmatrix}.
\end{equation}

Therefore, we have
\begin{equation}
  \{\mathbf{F}_{J,s}^H\mathbf{F}_{J,s}\mathbf{b}\mathbf{b}^H\}_{j,j} = \frac{1}{N}\sum_{n=1}^{N}\chi_{j,n}e^{-jk(r_{n,p}-r_{j,p})}.  
\end{equation}

Since only one submatrix is considered, we have
\begin{align}
& \chi_{j,n}e^{-jk(r_{n,p}-r_{j,p})} \overset{(a)}{\approx} \label{a1}\\
& \frac{1}{N} \sum_{z=(1-N)/2}^{(N-1)/2} e^{j\pi z (\sin\theta_j-\sin\theta_n)} e^{-jk(r_{n,p}-r_{j,p})}  \overset{j \neq n}{=} \label{a2}\\
&  e^{\frac{j\pi (\sin(\theta_j)-\sin(\theta_n))(1-N)}{2} -jk(r_{n,p}-r_{j,p})}\frac{1-e^{j\pi N (\sin(\theta_j)-\sin(\theta_n))}}{1-e^{j\pi (\sin(\theta_j)-\sin(\theta_n))}},\label{tx}
\end{align}
where the approximation (a) is obtained by performing the second-order Taylor Expansion to the distance term $r_{y,s,j}$  and $r_{y,s,n}$ 
for each $y=1, ..., N$ 
in the steering vector $\mathbf{b}(\theta_j,r_{s,j})$ 
and $\mathbf{b}(\theta_n,r_{s,n})$, which are
used in the calculation of $\chi_{j,n}$.
The second-order Taylor Expansion for $r_{y,s,j}$ around $r_{s,j}$ is expressed as 
\begin{equation}
r_{y,s,j}=r_{s,j} - t_y d \sin(\theta_j)+\frac{t_y^2d^2(1-\sin^2(\theta_j))}{r_{s,j}},
\end{equation}
where $t_y =  \frac{2y-N+1}{2}$. Besides, the transform from (\ref{a1}) to (\ref{a2}) also employs the following equation:
\begin{equation}
\frac{1-\sin^2(\theta_j)}{r_{s,j}} = \frac{1-\sin^2(\theta_n)}{r_{s,n}},
\end{equation}
which is a property of the joint angular-polar domain transform matrix.

Due to the definition of direction $\theta$ in (\ref{angleset}), $\sin(\theta_j)-\sin(\theta_n)$ is multiple of $2/N$. Therefore, based on (\ref{tx}), we have 
\begin{equation}
\chi_{j,n}e^{-jk(r_{n,p}-r_{j,p})} \approx \left\{
	\begin{aligned}
	0 \quad j \neq n,\\
	1 \quad j = n.\\
	\end{aligned}
	\right
	.
\end{equation}
Therefore, 
$\{\mathbf{F}_{J,s}^H\mathbf{F}_{J,s}\mathbf{b}\mathbf{b}^H\}_{j,j} \approx \frac{1}{N}$, and we have
\begin{equation}
{\rm Tr}(\mathbf{F}_{J,s}^H\mathbf{F}_{J,s}\mathbf{b}\mathbf{b}^H)\approx N*\frac{1}{N} =1.
\end{equation}
\vspace{-7mm}
\section{Proof of Lemma \ref{lemmamubound}}\label{app:lemma_mubound}

The term $\mu_{\mathbf{\Phi}}$ is expressed as $\mu_{\mathbf{\Phi}} = \max_{i\neq j} \vert\bm{\phi}_i^H\bm{\phi}_j\vert$, and we have $\bm{\phi}_i =  \mathbf{D}^{-1}\mathbf{W}\mathbf{f}_i$ and $\bm{\phi}_j = \mathbf{D}^{-1}\mathbf{W}\mathbf{f}_{j}$, where $\mathbf{f}_i$ and $\mathbf{f}_j$ are the $i$-th and $j$-th column of $\mathbf{F}_{J}$, respectively. Therefore, 
\begin{align}
& \vert\bm{\phi}_i^H\bm{\phi}_j\vert = \vert (\mathbf{D}^{-1}\mathbf{W}\mathbf{f}_i)^H \mathbf{D}^{-1}\mathbf{W}\mathbf{f}_j\vert = \nonumber\\
& \vert \mathbf{f}_i^H\mathbf{W}^H (\mathbf{D}^H\mathbf{D})^{-1}\mathbf{W}\mathbf{f}_j\vert = \vert \mathbf{W}^H\mathbf{f}_i^H \mathbf{C}^{-1}\mathbf{W}\mathbf{f}_j\vert = \nonumber\\
& \vert \mathbf{f}_i^H\mathbf{W}^H (\mathbb{B}\left\{ \mathbf{W}_1\mathbf{W}_1^H, \mathbf{W}_2\mathbf{W}_2^H, ... \mathbf{W}_Q\mathbf{W}_Q^H\right \})^{-1}\mathbf{W}\mathbf{f}_j \vert \nonumber \\
& = \vert \mathbf{f}_i^H\mathbf{W}^H (\mathbb{B}\left\{ (\mathbf{W}_1\mathbf{W}_1^H)^{-1}, ... (\mathbf{W}_Q\mathbf{W}_Q^H)^{-1}\right \})\mathbf{W}\mathbf{f}_j \vert \nonumber \\
& =  \vert \mathbf{f}_i^H\sum_{q=1}^{Q}\left\{ \mathbf{W}_q^H (\mathbf{W}_q\mathbf{W}_q^H)^{-1}\mathbf{W}_q\right \}\mathbf{f}_j \vert.
\end{align}
Considering that the beamforming is randomly generated, we further analyze the expectation of $\mu_{\mathbf{\Phi}}$: 
\vspace{-3mm}
\begin{align}
    & \mathbb{E}\left\{ \mu_{\mathbf{\Phi}} \right\} = \vert \mathbf{f}_i^H \mathbb{E}\left\{\sum_{q=1}^{Q}\left\{ \mathbf{W}_q^H (\mathbf{W}_q\mathbf{W}_q^H)^{-1}\mathbf{W}_q\right \} \right \}\mathbf{f}_j \vert \nonumber \\
    & = Q \vert \mathbf{f}_i^H \mathbb{E}\left\{ \mathbf{W}_q^H (\mathbf{W}_q\mathbf{W}_q^H)^ {-1}\mathbf{W}_q\right \} \mathbf{f}_j \vert, \forall q.
\end{align}
Considering the expectation of the $(x,y)$-th element of $\mathbf{W}_q\mathbf{W}_q^H$:
\vspace{-3mm}
\begin{equation}
\mathbb{E}\left\{(\mathbf{W}_q\mathbf{W}_q^H)_{x,y}\right\} = \mathbb{E}\left\{\sum_{z=1}^{N}((\mathbf{W}_q)_{x,z}(\mathbf{W}_q^H)_{z,y})\right\}.
\end{equation}
Because the modulus of each element of the beamforming matrix $\mathbf{W}_q$ is $1$, and each element is independently and randomly chosen with equal probability, we have
\begin{equation}
\mathbb{E}\left\{(\mathbf{W}_q\mathbf{W}_q^H)_{x,y}\right\} =
\begin{cases}
N,  & \text{if $x = y$}, \\
0, & \text{if $x \neq y$}.
\end{cases}
\end{equation}

Identically, considering the expectation of the $(x,y)$-th element of $\mathbf{W}_q^H\mathbf{W}_q$, we have 
\begin{equation}
\mathbb{E}\left\{(\mathbf{W}_q^H\mathbf{W}_q)_{x,y}\right\} =
\begin{cases}
N_{RF},  & \text{if $x = y$}, \\
0, & \text{if $x \neq y$}.
\end{cases}
\end{equation}
Therefore, we have $\mathbb{E}\left\{ \mathbf{W}_q^H (\mathbf{W}_q\mathbf{W}_q^H)^{-1}\mathbf{W}_q\right \} = \frac{N_{RF}}{N}\mathbf{I}_{N \times N}.$   

Since the modulus of each element of $\mathbf{f}_i$ is $\frac{1}{\sqrt{N}}, \forall i$ , we have 
\begin{equation}
\mathbb{E}\left\{ \mu_{\mathbf{\Phi}} \right\} = \frac{QN_{RF}}{N} \max_{i \neq j}\vert \mathbf{f}_i^H\mathbf{f}_j\vert < \frac{QN_{RF}}{N}.
\end{equation}

\section{Proof of Proposition \ref{prop:crlb}}\label{app:crlb}
For an arbitrary matrix $\mathbf{A}$ and an arbitrary vector $\mathbf{x}$, 
\begin{equation}
    \Vert\mathbf{A}\mathbf{x}\Vert_2 \geq \sigma_{\min}(\mathbf{A})\Vert\mathbf{x}\Vert_2,
\end{equation}
where $\sigma_{\min}(\mathbf{A})$ is the minimum singular value of matrix $\mathbf{A}$. Hence, we have
\begin{align}
&\Vert \hat{\mathbf{h}}_{H} - \mathbf{h}_{H}\Vert_2^2 = \nonumber \\ 
&\Vert \mathbf{F}_{J}(\hat{\mathbf{h}}_{J,H} - \mathbf{h}_{J,H})\Vert_2^2 \geq (\sigma_{\min}(\mathbf{F}_{J}))^2 \Vert (\hat{\mathbf{h}}_{J,H} - \mathbf{h}_{J,H})\Vert_2^2.
\end{align}
Thus, we have
\begin{equation}
\mathbb{E}\left \{\Vert \hat{\mathbf{h}}_{H} - \mathbf{h}_{H}\Vert_2^2 \right\} \geq (\sigma_{\min}(\mathbf{F}_{J}))^2 \frac{\sigma^2 {\rm card}(\Gamma)}{1+\frac{{\rm card}(\Gamma)QN_{RF}}{N}}.
\end{equation}
\end{appendices}


\end{document}